\pdfoutput=1

\documentclass[12pt,a4paper]{article}

\usepackage{ifthen} 
\usepackage{multirow}
\newboolean{pdflatex}
\setboolean{pdflatex}{true} 

\newboolean{articletitles}
\setboolean{articletitles}{true} 

\newboolean{uprightparticles}
\setboolean{uprightparticles}{false} 

\newboolean{inbibliography}
\setboolean{inbibliography}{false} 

\usepackage[top=1in, bottom=1.25in, left=1in, right=1in]{geometry}

\usepackage{microtype}
\usepackage{lineno}  
\usepackage{xspace} 
\usepackage{caption} 

\usepackage{graphicx}  
\usepackage{color}
\usepackage{colortbl}
\graphicspath{{./figs/}} 

\usepackage{amsmath,bm} 
\usepackage{amssymb}
\usepackage{amsfonts}
\usepackage{upgreek} 

\newcommand*\patchAmsMathEnvironmentForLineno[1]{%
\expandafter\let\csname old#1\expandafter\endcsname\csname #1\endcsname
\expandafter\let\csname oldend#1\expandafter\endcsname\csname
end#1\endcsname
 \renewenvironment{#1}%
   {\linenomath\csname old#1\endcsname}%
   {\csname oldend#1\endcsname\endlinenomath}%
}
\newcommand*\patchBothAmsMathEnvironmentsForLineno[1]{%
  \patchAmsMathEnvironmentForLineno{#1}%
  \patchAmsMathEnvironmentForLineno{#1*}%
}
\AtBeginDocument{%
\patchBothAmsMathEnvironmentsForLineno{equation}%
\patchBothAmsMathEnvironmentsForLineno{align}%
\patchBothAmsMathEnvironmentsForLineno{flalign}%
\patchBothAmsMathEnvironmentsForLineno{alignat}%
\patchBothAmsMathEnvironmentsForLineno{gather}%
\patchBothAmsMathEnvironmentsForLineno{multline}%
\patchBothAmsMathEnvironmentsForLineno{eqnarray}%
}

\usepackage{hyperref}    
\usepackage[all]{hypcap} 

\usepackage{xspace} 
\usepackage{upgreek}

\def\lhcb {\mbox{LHCb}\xspace}

\def\belle  {\mbox{Belle}\xspace}

\def\MagUp {\mbox{\em Mag\kern -0.05em Up}\xspace}

\ifthenelse{\boolean{uprightparticles}}%
{

 \def\Pmu         {\ensuremath{\upmu}\xspace}

 \def\Ppi         {\ensuremath{\uppi}\xspace}

 \def\Ppsi        {\ensuremath{\uppsi}\xspace}

 \def\PDelta      {\ensuremath{\Delta}\xspace}                 
 \def\PXi      {\ensuremath{\Xi}\xspace}                 
 \def\PLambda      {\ensuremath{\Lambda}\xspace}                 
 \def\PSigma      {\ensuremath{\Sigma}\xspace}                 
 \def\POmega      {\ensuremath{\Omega}\xspace}                 
 \def\PUpsilon      {\ensuremath{\Upsilon}\xspace}                 
 
 \def\PB      {\ensuremath{\mathrm{B}}\xspace}                 
                  
 \def\PD      {\ensuremath{\mathrm{D}}\xspace}

 \def\PJ      {\ensuremath{\mathrm{J}}\xspace}                 
 \def\PK      {\ensuremath{\mathrm{K}}\xspace}

 \def\Pb      {\ensuremath{\mathrm{b}}\xspace}                 
 \def\Pc      {\ensuremath{\mathrm{c}}\xspace}                 
                  
 \def\Pe      {\ensuremath{\mathrm{e}}\xspace}

 \def\Pi      {\ensuremath{\mathrm{i}}\xspace}

 \def\Ps      {\ensuremath{\mathrm{s}}\xspace}

}
{

 \def\Pmu         {\ensuremath{\mu}\xspace}

 \def\Ppi         {\ensuremath{\pi}\xspace}

 \def\Ppsi        {\ensuremath{\psi}\xspace}                 
                  
 \mathchardef\PDelta="7101
 \mathchardef\PXi="7104
 \mathchardef\PLambda="7103
 \mathchardef\PSigma="7106
 \mathchardef\POmega="710A
 \mathchardef\PUpsilon="7107
                  
 \def\PB      {\ensuremath{B}\xspace}                 
                  
 \def\PD      {\ensuremath{D}\xspace}

 \def\PJ      {\ensuremath{J}\xspace}                 
 \def\PK      {\ensuremath{K}\xspace}

 \def\Pb      {\ensuremath{b}\xspace}                 
 \def\Pc      {\ensuremath{c}\xspace}                 
                  
 \def\Pe      {\ensuremath{e}\xspace}

 \def\Pi      {\ensuremath{i}\xspace}

 \def\Ps      {\ensuremath{s}\xspace}

}

\makeatletter
\ifcase \@ptsize \relax
  \newcommand{\miniscule}{\@setfontsize\miniscule{4}{5}}
\or
  \newcommand{\miniscule}{\@setfontsize\miniscule{5}{6}}
\or
  \newcommand{\miniscule}{\@setfontsize\miniscule{5}{6}}
\fi
\makeatother

\DeclareRobustCommand{\optbar}[1]{\shortstack{{\miniscule (\rule[.5ex]{1.25em}{.18mm})}
  \\ [-.7ex] $#1$}}

\def\en         {{\ensuremath{\Pe^-}}\xspace}   
\def\ep         {{\ensuremath{\Pe^+}}\xspace}
 
\def\epem       {{\ensuremath{\Pe^+\Pe^-}}\xspace}

\def\mup        {{\ensuremath{\Pmu^+}}\xspace}
\def\mun        {{\ensuremath{\Pmu^-}}\xspace} 
\def\mumu       {{\ensuremath{\Pmu^+\Pmu^-}}\xspace}

\def\ellm       {{\ensuremath{\ell^-}}\xspace}
\def\ellp       {{\ensuremath{\ell^+}}\xspace}

\def\squark    {{\ensuremath{\Ps}}\xspace}

\def\cquark    {{\ensuremath{\Pc}}\xspace}

\def\bquark    {{\ensuremath{\Pb}}\xspace}

\def\pion   {{\ensuremath{\Ppi}}\xspace}

\def\pip    {{\ensuremath{\pion^+}}\xspace}

\def\kaon    {{\ensuremath{\PK}}\xspace}
  \def\Kbar    {{\kern 0.2em\overline{\kern -0.2em \PK}{}}\xspace}

\def\KorKbar    {\kern 0.18em\optbar{\kern -0.18em K}{}\xspace}

\def\Kp      {{\ensuremath{\kaon^+}}\xspace}

\def\Kstarz  {{\ensuremath{\kaon^{*0}}}\xspace}

\def\Kstar   {{\ensuremath{\kaon^*}}\xspace}

  \def\Dbar    {{\kern 0.2em\overline{\kern -0.2em \PD}{}}\xspace}
\def\D       {{\ensuremath{\PD}}\xspace}

\def\DorDbar    {\kern 0.18em\optbar{\kern -0.18em D}{}\xspace}
\def\Dz      {{\ensuremath{\D^0}}\xspace}

\def\B       {{\ensuremath{\PB}}\xspace}
\def\Bbar    {{\ensuremath{\kern 0.18em\overline{\kern -0.18em \PB}{}}}\xspace}

\def\BorBbar    {\kern 0.18em\optbar{\kern -0.18em B}{}\xspace}

\def\Bu      {{\ensuremath{\B^+}}\xspace}
\def\Bub     {{\ensuremath{\B^-}}\xspace}
\def\Bp      {{\ensuremath{\Bu}}\xspace}
\def\Bm      {{\ensuremath{\Bub}}\xspace}

\def\Bd      {{\ensuremath{\B^0}}\xspace}

\def\Bc      {{\ensuremath{\B_\cquark^+}}\xspace}

\def\jpsi     {{\ensuremath{{\PJ\mskip -3mu/\mskip -2mu\Ppsi\mskip 2mu}}}\xspace}
\def\psitwos  {{\ensuremath{\Ppsi{(2S)}}}\xspace}

  \def\Y#1S{\ensuremath{\PUpsilon{(#1S)}}\xspace}

\def\Lbar        {{\ensuremath{\kern 0.1em\overline{\kern -0.1em\PLambda}}}\xspace}
\def\LorLbar    {\kern 0.18em\optbar{\kern -0.18em \PLambda}{}\xspace}

\def\BF         {{\ensuremath{\mathcal{B}}}\xspace}

\newcommand{\decay}[2]{\ensuremath{#1\!\to #2}\xspace}         

\def\to                 {\ensuremath{\rightarrow}\xspace}

\def\qsq       {{\ensuremath{q^2}}\xspace}

\def\CP                {{\ensuremath{C\!P}}\xspace}

\def\AT#1     {\ensuremath{A_{\mathrm{T}}^{#1}}\xspace}           

\def\C#1      {\ensuremath{\mathcal{C}_{#1}}\xspace}                       
\def\Cp#1     {\ensuremath{\mathcal{C}_{#1}^{'}}\xspace}                    
\def\Ceff#1   {\ensuremath{\mathcal{C}_{#1}^{\mathrm{(eff)}}}\xspace}        
\def\Cpeff#1  {\ensuremath{\mathcal{C}_{#1}^{'\mathrm{(eff)}}}\xspace}       
\def\Ope#1    {\ensuremath{\mathcal{O}_{#1}}\xspace}                       
\def\Opep#1   {\ensuremath{\mathcal{O}_{#1}^{'}}\xspace}                    

\newcommand{\tev}{\ifthenelse{\boolean{inbibliography}}{\ensuremath{~T\kern -0.05em eV}\xspace}{\ensuremath{\mathrm{\,Te\kern -0.1em V}}}\xspace}
\newcommand{\gev}{\ensuremath{\mathrm{\,Ge\kern -0.1em V}}\xspace}
\newcommand{\mev}{\ensuremath{\mathrm{\,Me\kern -0.1em V}}\xspace}
\newcommand{\kev}{\ensuremath{\mathrm{\,ke\kern -0.1em V}}\xspace}
\newcommand{\ev}{\ensuremath{\mathrm{\,e\kern -0.1em V}}\xspace}
\newcommand{\gevc}{\ensuremath{{\mathrm{\,Ge\kern -0.1em V\!/}c}}\xspace}
\newcommand{\mevc}{\ensuremath{{\mathrm{\,Me\kern -0.1em V\!/}c}}\xspace}
\newcommand{\gevcc}{\ensuremath{{\mathrm{\,Ge\kern -0.1em V\!/}c^2}}\xspace}
\newcommand{\gevgevcccc}{\ensuremath{{\mathrm{\,Ge\kern -0.1em V^2\!/}c^4}}\xspace}
\newcommand{\mevcc}{\ensuremath{{\mathrm{\,Me\kern -0.1em V\!/}c^2}}\xspace}

\def\mum  {\ensuremath{{\,\upmu\mathrm{m}}}\xspace}

\def\invfb   {\ensuremath{\mbox{\,fb}^{-1}}\xspace}

\newcommand{\stat}{\ensuremath{\mathrm{\,(stat)}}\xspace}
\newcommand{\syst}{\ensuremath{\mathrm{\,(syst)}}\xspace}

\newcommand{\chisq}{\ensuremath{\chi^2}\xspace}

\def\deriv {\ensuremath{\mathrm{d}}}

\def\gsim{{~\raise.15em\hbox{$>$}\kern-.85em
          \lower.35em\hbox{$\sim$}~}\xspace}
\def\lsim{{~\raise.15em\hbox{$<$}\kern-.85em
          \lower.35em\hbox{$\sim$}~}\xspace}

\def\ptot       {\mbox{$p$}\xspace}
\def\pt         {\mbox{$p_{\mathrm{ T}}$}\xspace}

\def\mrad{\ensuremath{\mathrm{ \,mrad}}\xspace}

\def\evtgen     {\mbox{\textsc{EvtGen}}\xspace}

\def\geant      {\mbox{\textsc{Geant4}}\xspace}

\def\photos     {\mbox{\textsc{Photos}}\xspace}

\def\pythia     {\mbox{\textsc{Pythia}}\xspace}

\def\tell1  {TELL1\xspace}
\def\ukl1   {UKL1\xspace}

\newcommand{\eg}{\mbox{\itshape e.g.}\xspace}
\newcommand{\ie}{\mbox{\itshape i.e.}\xspace}

\def\btosll               {\ensuremath{\decay{\bquark}{\squark\ellp\ellm}}\xspace}

\def\BuToKmm      {\decay{\Bu}{\Kp\mumu}}

\def\BuToKmm      {\decay{\Bu}{\Kp\mumu}}

\def\mkmm  {\ensuremath{m_{K\mu\mu}}\xspace}

\def\Cnine {\ensuremath{\mathcal{C}_{9}}\xspace}

\def\BuJpsiK       {\decay{\Bp}{ \jpsi \Kp}}
\def\BuKMuMu     {\decay{\Bp}{ \Kp \mup\mun}}

\usepackage{cite} 
\usepackage{mciteplus}

\makeatletter
\DeclareRobustCommand*{\bfseries}{%
  \not@math@alphabet\bfseries\mathbf
  \fontseries\bfdefault\selectfont
  \boldmath
}
\makeatother

\usepackage{longtable} 

\def\BuJpsiK       {\decay{\Bp}{ \jpsi \Kp}}
\def\BuKMuMu     {\decay{\Bp}{ \Kp \mup\mun}}

\begin{document}

\renewcommand{\thefootnote}{\fnsymbol{footnote}}
\setcounter{footnote}{1}

\begin{titlepage}
  \pagenumbering{roman}

  \vspace*{-1.5cm}
  \centerline{\large EUROPEAN ORGANIZATION FOR NUCLEAR RESEARCH (CERN)}
  \vspace*{1.5cm}
  \noindent
  \begin{tabular*}{\linewidth}{lc@{\extracolsep{\fill}}r@{\extracolsep{0pt}}}
    \ifthenelse{\boolean{pdflatex}}
    {\vspace*{-2.7cm}\mbox{\!\!\!\includegraphics[width=.14\textwidth]{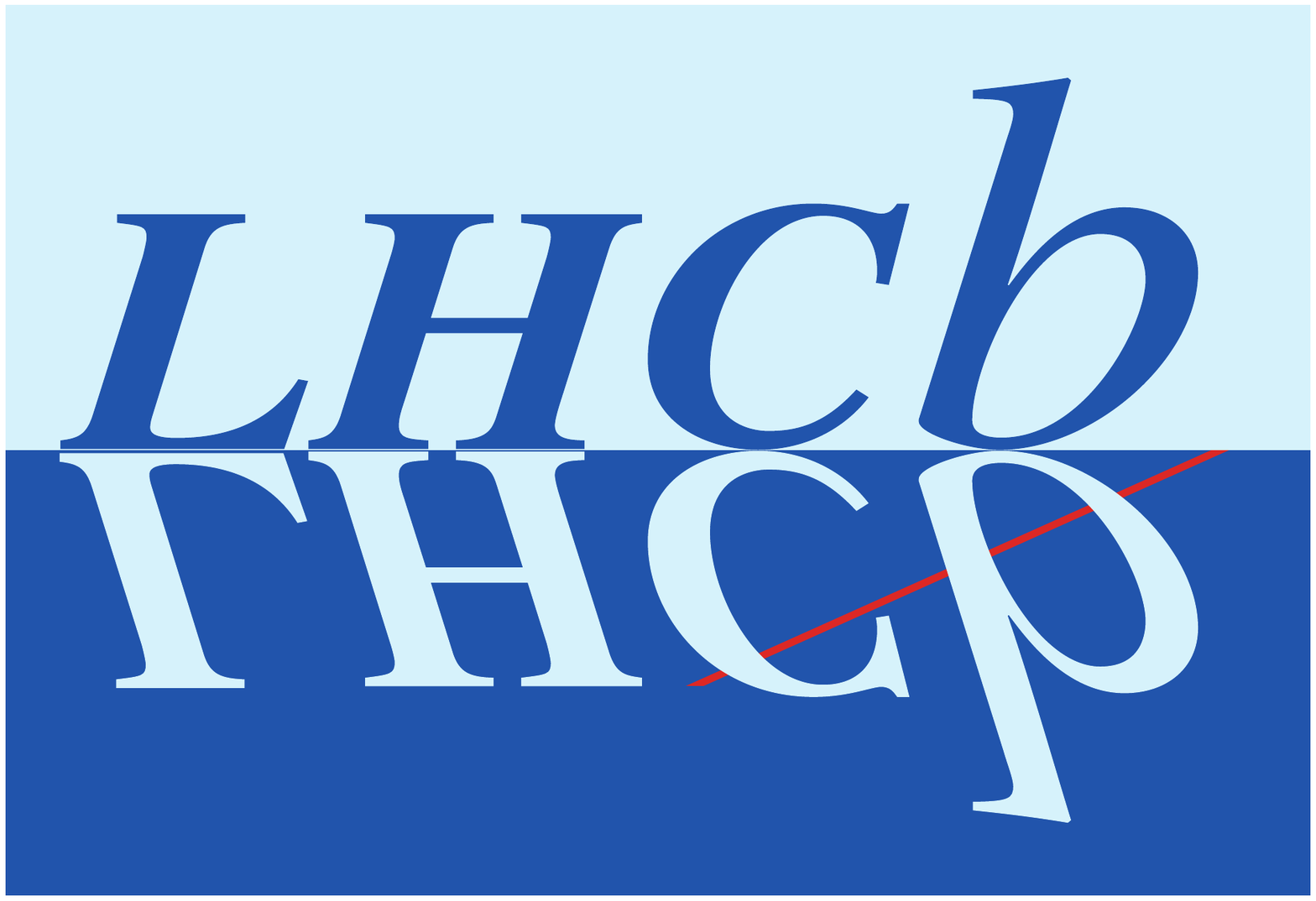}} & &}%
    {\vspace*{-1.2cm}\mbox{\!\!\!\includegraphics[width=.12\textwidth]{lhcb-logo.eps}} & &}%
    \\
    & & CERN-EP-2016-301 \\  
    & & LHCb-PAPER-2016-045 \\  
    & & \today \\ 
    & & \\
  \end{tabular*}

  \vspace*{0.9cm}

  {\normalfont\bfseries\boldmath\LARGE
    \begin{center}
  Measurement of the phase difference between short- and long-distance amplitudes in the \BuToKmm decay
    \end{center}
  }

  \vspace*{0.90cm}

  \begin{center}
    The LHCb collaboration\footnote{Authors are listed at the end of this paper.}
  \end{center}

  \vspace{\fill}

  \begin{abstract}
    \noindent
    A measurement of the phase difference between the short- and long-distance contributions to the \BuToKmm decay is performed by analysing the dimuon mass distribution.  
    The analysis is based on $pp$ collision data corresponding to an integrated luminosity of 3\invfb collected by the \lhcb experiment in 2011 and 2012. 
    The long-distance contribution  to the \BuToKmm decay is modelled as a sum of relativistic Breit--Wigner amplitudes representing different vector meson resonances decaying to muon pairs, each with their own magnitude and phase. 
    The measured phases of the \jpsi and \psitwos resonances are such that the interference with the short-distance component in dimuon mass regions far from their pole masses is small. 
    In addition, constraints are placed on the Wilson coefficients, \Cnine and $\mathcal{C}_{10}$, and the branching fraction of the short-distance component is measured.
  \end{abstract}

  \vspace*{4.0cm}

  \begin{center}
    Published in  Eur. Phys. J. C
  \end{center}

  \vspace{\fill}

  {\footnotesize 
    \centerline{\copyright~CERN on behalf of the \lhcb collaboration, licence \href{http://creativecommons.org/licenses/by/4.0/}{CC-BY-4.0}.}}
    \vspace*{2mm}

  \end{titlepage}


  \newpage
  \setcounter{page}{2}
  \mbox{~}
%
%
%
%

  \cleardoublepage


\renewcommand{\thefootnote}{\arabic{footnote}}
\setcounter{footnote}{0}



\pagestyle{plain} 
\setcounter{page}{1}
\pagenumbering{arabic}


%

\clearpage

\section{Introduction}
\label{sec:Introduction}

The decay \BuToKmm receives contributions from short-distance \btosll flavour-changing neutral-current (FCNC) transitions and  long-distance contributions from intermediate hadronic resonances.
In the Standard Model (SM), FCNC transitions are forbidden at tree level and must occur via a loop-level process. 
In many extensions of the SM, new particles can contribute to the amplitude of the \btosll process changing the rate of the decay or the distribution of the final-state particles.
Decays like \BuToKmm are therefore sensitive probes of physics beyond the SM.

Recent global analyses of measurements involving \btosll processes report deviations from SM predictions at the level of four standard deviations~\cite{
Descotes-Genon:2013wba,
Altmannshofer:2013foa,
Altmannshofer:2014cfa,
Mahmoudi:2014mja,
Crivellin:2015mga,
Descotes-Genon:2015uva,
Hurth:2016fbr,
Jager:2012uw,
Beaujean:2013soa,
Hurth:2013ssa,
Gauld:2013qja,
Datta:2013kja,
Lyon:2014hpa,
Descotes-Genon:2014uoa,
AltAndStraubLatest}.
These differences could be explained by new short-distance contributions from non-SM particles~\cite{Descotes-Genon:2013wba,Altmannshofer:2013foa,Altmannshofer:2014cfa,Crivellin:2015mga,Mahmoudi:2014mja,Datta:2013kja,Buttazzo:2016kid} or could indicate a problem with existing SM predictions~\cite{Ciuchini:2015qxb,AltAndStraubLatest,Lyon:2014hpa}. 
To explain the observed tensions, long-distance effects would need to be sizeable in dimuon mass regions far from the pole masses of the resonances. 
Therefore, it is important to understand how well these long-distance effects are modelled in the SM and how they interfere with the short-distance contributions. 
Previous measurements of \btosll processes~\cite{Lees:2012tva,Wei:2009zv,Aaltonen:2011cn,CMSKstmm,LHCb-PAPER-2014-006,LHCb-PAPER-2015-051} excluded regions of dimuon mass around the $\phi$, \jpsi and \psitwos resonances. 
The amplitude in these mass regions is dominated by the narrow vector resonances and has a large theoretical uncertainty. 
These dimuon regions are therefore considered insensitive to new physics effects.

In this paper, a first measurement of the phase difference between the contributions to the short-distance and the narrow-resonance amplitudes in the $\BuToKmm$ decay is presented.\footnote{The inclusion of charge-conjugate processes is implied throughout.} 
For the first time, the branching fraction of the short-distance component is determined without interpolation across the \jpsi and \psitwos regions.
The measurement is performed through a fit to the full dimuon mass spectrum, $m_{\mu\mu}$, using a model describing the vector resonances as a sum of relativistic Breit--Wigner amplitudes. 
This approach is similar to that of Refs.~\cite{Kruger:1996cv,Lyon:2014hpa}, with the difference that the magnitudes and phases of the resonant amplitudes are  determined using the LHCb data rather than using the external information on the cross-section for \decay{\ep\en}{{\rm hadrons}}  from the BES collaboration~\cite{Bai:2001ct}. 
The model includes the $\rho$, $\omega$, $\phi$, \jpsi and \psitwos resonances, as well as broad charmonium states ($\psi(3770)$, $\psi(4040)$, $\psi(4160)$ and $\psi(4415)$) above the open charm threshold. 
Evidence for the $\psi(4160)$ resonance in the dimuon spectrum of \BuToKmm decays has been previously reported by LHCb in Ref.~\cite{LHCb-PAPER-2013-039}. 
The continuum of broad states with pole masses above the maximum $m_{\mu\mu}$ value allowed in the decay is neglected.

The measurement presented in this paper is performed using a data set corresponding to 3\invfb of integrated luminosity collected by the LHCb experiment in $pp$ collisions during 2011 and 2012 at $\sqrt{s}$ = 7\tev and 8\tev. 
The paper is organised as follows: Section~\ref{sec:detector} describes the LHCb detector and the procedure used to generate simulated events; 
the reconstruction and selection of $\BuToKmm$ decays are described in Sec.~\ref{sec:selection}; 
Section~\ref{sec:model} describes the $m_{\mu\mu}$ distribution of  $\BuToKmm$ decays, including the model for the various resonances appearing in the dimuon mass spectrum; 
the fit procedure to the dimuon mass spectrum, including the methods to correct for the detection and selection biases, is discussed in Sec.~\ref{sec:fit}. 
The results and associated systematic uncertainties are discussed in Secs.~\ref{sec:results} and~\ref{sec:systematics}.
Finally, conclusions are presented in Sec.~\ref{sec:conclusions}.

\section{Detector and simulation}
\label{sec:detector}

The \lhcb detector~\cite{Alves:2008zz,LHCb-DP-2014-002} is a single-arm forward
spectrometer, covering the \mbox{pseudorapidity} range $2<\eta <5$,
designed to study the production and decay of particles containing \bquark or \cquark quarks.
The detector includes a high-precision tracking system divided
into three subsystems: a silicon-strip vertex detector 
surrounding the $pp$ interaction region, a large-area silicon-strip detector that is located
upstream of a dipole magnet with a bending power of about
$4{\mathrm{\,Tm}}$, and three stations of silicon-strip detectors and straw
drift tubes situated downstream of the magnet.
The tracking system provides a measurement of the momentum, \ptot, of charged particles with
a relative uncertainty that varies from 0.5\% at low momentum to 1.0\% at 200\gevc.
The momentum scale of tracks in the data is calibrated using the \Bp and \jpsi masses measured in \decay{\Bp}{\jpsi\Kp} decays~\cite{LHCb-PAPER-2012-048}. 
The minimum distance of a track to a \mbox{primary vertex (PV)}, the impact parameter (IP), is measured with a resolution of $(15+29/\pt)\mum$,
where \pt is the component of the momentum transverse to the beam, in\,\gevc.
Different types of charged hadrons are distinguished using information
from two ring-imaging Cherenkov {\mbox{detectors (RICH)}}. 
Photons, electrons and hadrons are identified by a calorimeter system consisting of
scintillating-pad and preshower detectors, an electromagnetic
calorimeter and a hadronic calorimeter. Muons are identified by a
system composed of alternating layers of iron and multiwire
proportional chambers. 
The online event selection is performed by a trigger~\cite{LHCb-DP-2012-004}, 
which consists of a hardware stage, based on information from the calorimeter and muon
systems, followed by a software stage, which applies a full event
reconstruction.

A large sample of simulated events is used to determine the
effect of the detector geometry, trigger, and selection criteria
on the dimuon mass distribution of the \mbox{\BuToKmm} decay. 
In the simulation, $pp$ collisions are generated using
\pythia 8~\cite{Sjostrand:2006za,*Sjostrand:2007gs} with a specific
\lhcb configuration~\cite{LHCb-PROC-2010-056}.  The decay of the \Bp
meson is described by \evtgen~\cite{Lange:2001uf}, which generates
final-state radiation using
\photos~\cite{Golonka:2005pn}.  As described in
Ref.~\cite{LHCb-PROC-2011-006}, the \geant
toolkit~\cite{Allison:2006ve, *Agostinelli:2002hh} is used to
implement the interaction of the generated particles with the detector
and its response. Data-driven corrections
are applied to the simulation following the procedure of
Ref.~\cite{LHCb-PAPER-2015-051}. These corrections account for the small
level of mismodelling of the detector occupancy, the $\Bp$ momentum and
vertex quality, and the particle identification (PID)
performance.
The momentum of every reconstructed track in the simulation is also smeared by a small amount in order to better match the mass resolution of the data.

\section{Selection of signal candidates}
\label{sec:selection}

In the trigger for the 7\tev (8\tev) data, at least one of the muons is required to have $\pt>1.48\gevc$ ($\pt > 1.76\gevc$) and one of the final-state particles is required to have both $\pt>1.4\gevc$ ($\pt >1.6\gevc$) and an ${\rm IP} > 100\mum$ with respect to all PVs in the event; if this final-state particle is identified as a muon, $\pt > 1.0\gevc$ is required instead.  Finally, the tracks of two or more of the final-state particles are required to form a vertex that is significantly displaced from all PVs.

In the offline selection, signal candidates are built from a pair of oppositely tracks that are identified as muons. 
The muon pair is then combined with a charged track that is identified as a kaon by the RICH detectors. 
The signal candidates are required to pass a set of loose preselection requirements that are identical to those described in Ref.~\cite{LHCb-PAPER-2013-039}. 
These requirements exploit the decay topology of \BuToKmm transitions and restrict the data sample to candidates with good-quality vertex and track fits. 
Candidates are required to have a reconstructed $\Kp\mumu$ mass, $\mkmm$, in the range $5100<m_{K\mu\mu}<6500\mevcc$. 

Combinatorial background, where particles from different decays are mistakenly combined, is further suppressed with the use of a Boosted Decision Tree~(BDT)~\cite{Breiman,AdaBoost} using kinematic and geometric information. 
The BDT is identical to that described in Ref.~\cite{LHCb-PAPER-2013-039} and uses the same working point.
The efficiency of the BDT for signal is uniform with respect to \mkmm. 

Specific background processes can mimic the signal if their final states are misidentified or partially reconstructed.  
The requirements described in Ref.~\cite{LHCb-PAPER-2013-039} reduce the overall contribution of the background from such decay processes to a level of less than 1\% of the expected signal yield in the full mass region.  
The largest remaining specific background contribution comes from \decay{\Bp}{\pip\mumu} decays (including \decay{\Bp}{\jpsi\pip} and \decay{\Bp}{\psitwos\pip}), where the pion is mistakenly identified as a kaon. 

\begin{figure}[!htb]
\centering
\includegraphics[width=0.8\linewidth]{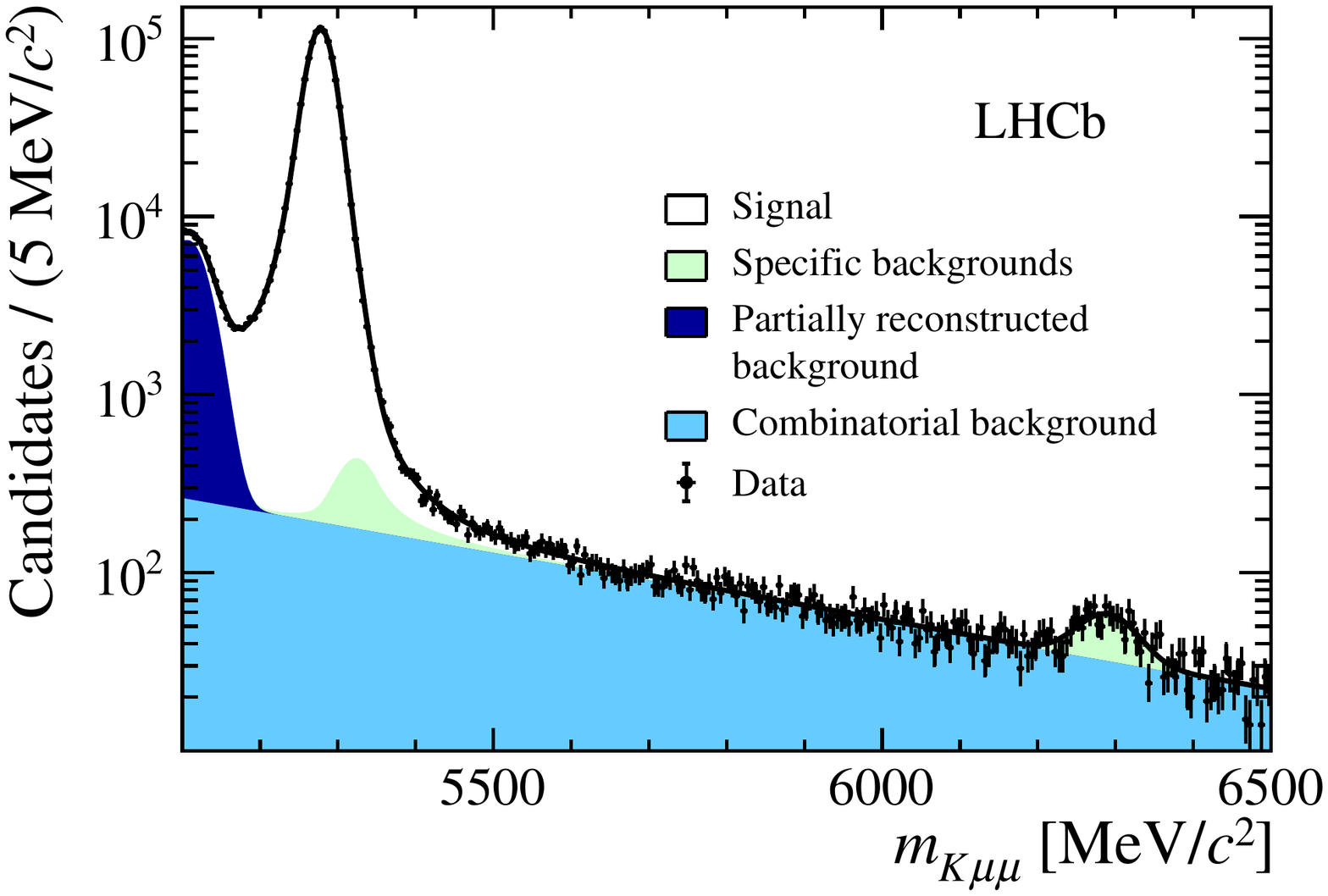}
\caption{
Reconstructed $\Kp\mumu$ mass of the selected \decay{\Bp}{\Kp\mumu} candidates. The fit to the data is described in the text. \label{fig:candidates} 
}

\end{figure}

The $\Kp\mumu$ mass of the selected candidates is shown in Fig.~\ref{fig:candidates}. 
The signal is modelled by the sum of two Gaussian functions and a Gaussian function with power-law tails on both sides of the peak; these all share a common peak position.
A Gaussian function is used to describe a small contribution from \Bc decays around the known \Bc mass~\cite{Olive:2016xmw}.
Combinatorial background is described by an exponential function with a negative gradient. 
At low $m_{K\mu\mu}$, the  background is dominated by partially reconstructed $b$-hadron decays,  \eg from \decay{B^{\{+,0\}}}{K^{*\{+,0\}}\mumu} decays in which the pion from the $K^{*\{+,0\}}$ is not reconstructed. 
This background component is modelled using the upper tail of a Gaussian function. 
The shape of the background from \decay{\Bp}{\pip\mumu} decays is taken from a sample of simulated events. 
Integrating the signal component in a $\pm40$\mevcc window about the known \Bp mass~\cite{Olive:2016xmw} yields 980\,000 \BuKMuMu decays.

When computing $m_{\mu\mu}$, a kinematic fit is performed to the selected candidates. 
In the fit, the $m_{K\mu\mu}$ mass is constrained to the known \Bp mass and the candidate is required to originate from one of the PVs in the event.
For simulated  \BuJpsiK decays, this improves the resolution in $m_{\mu\mu}$ by about a factor of two.

\section{Differential decay rate}
\label{sec:model}

Following the notation of Ref.~\cite{Bailey:2015dka}, the \CP-averaged differential decay rate of \mbox{\BuKMuMu} decays as a function of the dimuon mass squared, $\qsq \equiv m_{\mu\mu}^{2}$, is given by
\begin{align}
    \frac{\deriv\Gamma}{\deriv q^2} = & \frac{G_F^2\alpha^2|V_{tb}V_{ts}^*|^2}{128\pi^5} |\bm{k}|\beta
        \left\{ \vphantom{\left|2\mathcal{C}_7\frac{m_b+m_s}{m_B+m_K} f_T(q^2)\right|^2}
        \frac{2}{3}|\bm{k}|^2\beta^2 \left|\mathcal{C}_{10} f_+(q^2)\right|^2 \right.
    +
        \frac{4m_{\mu}^2 (m_B^2-m_K^2)^2 }{q^2 m_B^2}
        \left| \mathcal{C}_{10} f_0(q^2)\right|^2
    \nonumber \\ & { } + \left.
        |\bm{k}|^2 \left[1 - \frac{1}{3}\beta^2 \right]
        \left| \mathcal{C}_9 f_+(q^2) + 2\mathcal{C}_7\frac{m_b+m_s}{m_B+m_K} f_T(q^2) \right|^2
        \right\},
    \label{eq:dGamma}
\end{align}
where $|\bm{k}|$ is the kaon momentum in the \Bp meson rest frame.
Here $m_K$ and $m_B$ are the masses of the \Kp and \Bp mesons while $m_s$ and $m_b$ refer to the $s$ and $b$ quark masses as defined in Ref.~\cite{Bailey:2015dka}, $m_{\mu}$ is the muon mass and $\beta^2=1-4m_\mu^2/q^2$.
The constants $G_F$, $\alpha$, and $V_{tq}$ are the Fermi constant, the QED fine structure constant, and CKM matrix elements, respectively.  
The parameters $f_{0,+,T}$ denote the scalar, vector and tensor $B\to K$ form factors.
The $\mathcal{C}_i$ are the Wilson coefficients in an effective field theory description of the decay. 
The coefficient $\mathcal{C}_9$ corresponds to the coupling strength of the vector current operator, $\mathcal{C}_{10}$ to the axial-vector current operator  and $\mathcal{C}_7$ to the electromagnetic dipole operator. 
The operator definitions and the numerical values of the Wilson coefficients in the SM can be found in Ref.~\cite{Altmannshofer:2008dz}. 
Right-handed Wilson coefficients, conventionally denoted $\mathcal{C}'_i$, are suppressed in the SM and are ignored in this analysis. 
The Wilson coefficients $\mathcal{C}_9$ and $\mathcal{C}_{10}$ are assumed to be real. 
This implicitly assumes that there is no weak phase associated with the short-distance contribution. 
In general, $C\!P$-violating effects are expected to be small across the $m_{\mu\mu}$ range 
with the exception of the region around the $\rho$ and $\omega$ resonances, which
enter with different strong and weak phases~\cite{Alok:2011gv}. The small size of the $C\!P$ 
asymmetry between \Bm and \Bp decays is confirmed in Ref.~\cite{LHCb-PAPER-2014-032}. 
In the present analysis, there is no sensitivity to $C\!P$-violating effects at low masses and 
therefore the phases of the resonances are taken to be the same for \Bp and \Bm decays throughout. 

Vector resonances, which produce dimuon pairs via a virtual photon, mimic a contribution to $\mathcal{C}_9$.
These long-distance hadronic contributions to the \decay{\Bp}{\Kp\mumu} decay are taken into account by introducing an effective Wilson coefficient in place of $\mathcal{C}_{9}$ in Eq.~\ref{eq:dGamma},
\begin{equation}
\mathcal{C}_{9}^{\rm eff} = \mathcal{C}_{9}+Y(\qsq), 
\end{equation}
where the term $Y(q^2)$ describes the sum of resonant and continuum hadronic states appearing in the dimuon mass spectrum. 
In this analysis $Y(\qsq)$ is replaced by the sum of vector meson resonances $j$ such that 
\begin{equation}
\mathcal{C}_{9}^{\rm eff} = \mathcal{C}_{9} + \sum\limits_{j} \eta_{j}e^{i\delta_{j}} A_{j}^{\rm res}(\qsq), 
\label{eq:yqsq}
\end{equation}
where $\eta_{j}$ is the magnitude of the resonance amplitude and $\delta_j$ its  phase relative to $C_9$. These phase differences are one of the main results of this paper. 
The \qsq dependence of the magnitude and phase of the resonance is parameterised by $A_{j}^{\rm res}(\qsq)$. 
The resonances included are the $\omega$, $\rho^0$, $\phi$, $\jpsi$, $\psitwos$, $\psi(3770)$, $\psi(4040)$, $\psi(4160)$ and $\psi(4415)$. 
Contributions from other broad resonances and hadronic continuum states are ignored, as are contributions from weak annihilation~\cite{Feldmann:2002iw,Khodjamirian:2012rm,Lyon:2013gba}. 
No systematic uncertainties are attributed to these assumptions,
which are part of the model that defines the analysis framework
of this paper.

The function $A_{j}^{\rm res}(\qsq)$ is taken to have the form of a relativistic Breit--Wigner function for the $\omega$, $\rho^0$, $\phi$, $\jpsi$, $\psitwos$ and $\psi(4040)$, $\psi(4160)$ and $\psi(4415)$ resonances,
\begin{equation}
A_{j}^{\rm res}(\qsq) = \frac{m_{0\,j}\Gamma_{0\,j}}{(m_{0\,j}^{2}-\qsq) - i m_{0\,j}\Gamma_{j}(\qsq)},
\label{eq:BW}
\end{equation}
where $m_{0\,j}$ is the pole mass of the $j^{\rm th}$ resonance and $\Gamma_{0\,j}$ its natural width. 
The running width $\Gamma_{j}(q^{2})$ is given by
\begin{equation}
\Gamma_{j}(\qsq) = \frac{p}{p_{0\,j}}\frac{m_{0\,j}}{\sqrt{\qsq}}\Gamma_{0\,j},
\label{eq:BWwidth}
\end{equation}
where $p$ is the momentum of the muons in the rest frame of the dimuon system evaluated at $q$, and $p_{0\,j}$ is the momentum evaluated at the mass of the resonance.
To account for the open charm threshold, the lineshape of the $\psi(3770)$ resonance is described by a Flatt\'{e}  function~\cite{Flatte:1976xu} with a width defined as
\begin{equation}
\Gamma_{\psi(3770)}(\qsq) = \frac{p}{p_{0\,j}}\frac{m_{0\,j}}{\sqrt{\qsq}}\left[\Gamma_{1} + \Gamma_{2}\sqrt{\frac{1-(4m_{D}^{2}/\qsq)}{1-(4m_{D}^{2}/q^2_{0})}} \right]\,,
\label{eq:flatte}
\end{equation}
where $m_D$ is the mass of the \Dz meson and $q^2_{0}$ is the \qsq value at the pole mass of the $\psi(3770)$. The coefficients $\Gamma_{1}=0.3\mevcc$ and $\Gamma_{2} = 27\mevcc$ are taken from Ref.~\cite{Olive:2016xmw} and correspond to the sum of the partial widths of the $\psi(3770)$ to states below and above the open charm threshold. For $\qsq < 4 m_D^2$, the phase-space factor accompanying $\Gamma_2$ in Eq.~\ref{eq:flatte} becomes complex.

The form factors are parameterised according to Ref.~\cite{Bourrely:2008za} as
\begin{align}
f_0(\qsq) & =  \frac{1}{1 - \qsq/m_{B_{s0}^*}^2}  \sum\limits_{i=0}^{N-1} b^{0}_i z^i \,, \\
f_{+,T}(\qsq) & = \frac{1}{1 - \qsq/m_{B_s^*}^2} \sum\limits_{i=0}^{N-1} b^{+,T}_i \left[ z^i - (-1)^{i - N} \left(\frac{i}{N}\right) z^{N} \right] \, ,
\label{eq:FF}
\end{align} 
with, for this analysis, $N=3$.
Here $m_{B_s^*} (m_{B_{s0}^*})$ is the mass of the lowest-lying excited $B_s$ meson with $J^P=1^- (0^+)$. 
The coefficients $b^{+}_{i}$ are allowed to vary in the fit to the data subject to constraints 
from Ref.~\cite{Bailey:2015dka}, whereas the coefficients $b^{0}_{i}$ and $b^{T}_{i}$ are fixed to 
their central values. The function $z$ is defined by the mapping
\begin{align}
z(\qsq) \equiv \frac{\sqrt{t_+ - \qsq} - \sqrt{t_+ - t_0}}{\sqrt{t_+ - \qsq} + \sqrt{t_+ - t_0}}
\end{align} 
with 
\begin{align} 
t_+  \equiv (m_B - m_K)^2 
\end{align}
and
\begin{align}
t_0 \equiv (m_B + m_K)(\sqrt{m_B} - \sqrt{m_K})^2~.
\end{align}

\section{Fit to the $m_{\mu\mu}$ distribution}
\label{sec:fit}

In order to determine the magnitudes and phases of the different resonant contributions, a maximum likelihood fit in 538 bins is performed to the distribution of the reconstructed dimuon mass, $m_{\mu\mu}^{\rm rec}$, of candidates with $m_{K\mu\mu}$ in a $\pm40$\mevcc window about the known \Bp mass. 
The $m^{\rm rec}_{\mu\mu}$ distribution of the \decay{\Bp}{\Kp\mumu} decay is described by  
\begin{equation}
R( m_{\mu\mu}^{\rm rec}, m_{\mu\mu} ) \otimes \left( \varepsilon(m_{\mu\mu})  \frac{\deriv \Gamma}{\deriv \qsq} \frac{\deriv\qsq}{\deriv m_{\mu\mu}} \right)\;,
\label{eq:fullmodel}
\end{equation} 
\ie by Eq.~\ref{eq:dGamma}, multiplied by the detector efficiency, $\varepsilon$, as a function of the true dimuon mass, $m_{\mu\mu}$, and convolved with the experimental mass resolution $R$ discussed in Sec.~\ref{sec:resolution}.

\subsection{Signal model}

The magnitudes and phases of the resonances are allowed to vary in the fit, as are the Wilson coefficients $\mathcal{C}_9$ and $\mathcal{C}_{10}$. 
As the contribution of $\mathcal{C}_7$ to the total decay rate is small, it is fixed to its SM value of  \mbox{$\mathcal{C}_7^{\rm SM} = -0.304\pm0.006$}~\cite{Altmannshofer:2008dz}. 

 The form factor $f_{+}(\qsq)$ is constrained in the fit according to its value and uncertainty from Ref.~\cite{Bailey:2015dka}. 
The form factors $f_{0}(\qsq)$ and $f_{T}(\qsq)$ have a limited impact on the normalisation and shape of Eq.~\ref{eq:dGamma}, and are fixed to their values from Ref.~\cite{Bailey:2015dka}.
The masses and widths of the broad resonances above the open charm threshold are constrained according to their values in Ref.~\cite{Ablikim:2007gd}. 
The masses and widths of the $\rho$, $\omega$ and $\phi$ mesons and the widths of the \jpsi and \psitwos mesons are fixed to their known values~\cite{Olive:2016xmw}.  
The large magnitude of the \jpsi and \psitwos amplitudes makes the fit very sensitive to the position of the pole mass of these resonances. 
Due to some residual uncertainty on the momentum scale in the data, the pole masses of the \jpsi and \psitwos mesons are allowed to vary in the fit. 

The short-distance component is normalised to the branching fraction of \mbox{\decay{\Bp}{\jpsi\Kp}} measured by the $B$-factory experiments~\cite{Olive:2016xmw}. After correcting for isospin asymmetries in the production of the \Bp mesons at the $\Upsilon(4S)$, the branching fraction is \mbox{$\BF(\decay{\Bp}{\jpsi\Kp})=(9.95\pm0.32)\times 10^{-4}$}~\cite{Jung:2015yma}.
This is further multiplied by \mbox{$\BF(\decay{\jpsi}{\mumu}) = (5.96 \pm 0.03)\times 10^{-2}$}~\cite{Olive:2016xmw} to account for the decay of the \jpsi meson. 
The branching fraction of the decay \decay{\Bp}{\Kp\mumu} via an intermediate resonance $j$ is computed from the fit as
\begin{align}
\tau_{B}
\frac{G_F^2\alpha^2|V_{tb}V_{ts}^*|^2}{128\pi^5} \int\limits_{4 m^2_\mu}^{(m_B - m_K)^{2}} 
        |\bm{k}|^3 \left[ \beta - \frac{1}{3}\beta^3 \right]
        \left| f_+(q^2) \right|^2 \left| \eta_j \right|^2 \left| A_j^{\rm res} (\qsq) \right|^2 \deriv\qsq\,,
    \label{eq:br:constraint}
\end{align}
where $\tau_B$ is the lifetime of the \Bp meson.
The branching fractions of \decay{\Bp}{\rho\Kp}, \decay{\Bp}{\omega\Kp}, \decay{\Bp}{\phi\Kp} and \decay{\Bp}{\psi(3770)\Kp} are also constrained assuming factorisation between the \B decay and the subsequent decay of the intermediate resonance to a muon pair. 
These branching fractions are taken from Ref.~\cite{Olive:2016xmw}.

\subsection{Mass resolution}
\label{sec:resolution}

The convolution of the resolution function with the signal model is implemented using a fast Fourier transform technique~\cite{Cooley:1965zz,FFTW05}. 
The fit to the data is performed in three separate regions of dimuon mass: $300 \leq m_{\mu\mu}^{\rm rec} \leq 1800\mevcc$, $1800 < m_{\mu\mu}^{\rm rec}  \leq 3400\mevcc$ and $3400 < m_{\mu\mu}^{\rm rec} \leq 4700\mevcc$. 

To increase the speed of the fit, the resolution is treated as constant within these regions using the resolution at the $\phi$, \jpsi and \psitwos pole masses.
The impact of this assumption on the measured phases of the \jpsi and \psitwos resonances has been tested using pseudoexperiments and found to be negligible. 
This is to be expected as the spectra in all other regions vary slowly in comparison to the resolution function.
The resolution is modelled using the sum of a Gaussian function, $G$, and a Gaussian function with power-law tails on the lower and upper side of the peak, $C$,
\begin{align}
\begin{split}
R(m^{\rm rec}_{\mu\mu},m_{\mu\mu}) = f\,& G  ( m^{\rm rec}_{\mu\mu}, m_{\mu\mu}, \sigma_G )  + \\
&  (1 - f ) \, C( m^{\rm rec}_{\mu\mu}, m_{\mu\mu}, \sigma_C, n_{\rm l}, n_{\rm u}, \alpha_{\rm l}, \alpha_{\rm u} ) \,.
\end{split}
\end{align}
The component with power-law tails is defined as 
\begin{equation}
\begin{split}
 C( m^{\rm rec}_{\mu\mu}, m_{\mu\mu}, \sigma_C, n_{\rm l}, n_{\rm u}, \alpha_{\rm l}, \alpha_{\rm u} ) & \propto  
\left\{
\begin{array}{ll} 
A_{\rm l} \, \left( B_{\rm l} - \delta \right)^{-n_{\rm l}} & ~\text{if}~\delta < \alpha_{\rm l} \phantom{\frac{1}{1}} \\
{\rm exp}(-\delta^2/2) & ~\text{if}~\alpha_{\rm l} < \delta < \alpha_{\rm u}  \\
A_{\rm u} \, \left( B_{\rm u} + \delta \right)^{-n_{\rm u}} & ~\text{if}~\delta > \alpha_{\rm u} \\
\end{array} \right. \,,
\end{split}
\end{equation} 
with 
\begin{align} 
\begin{split}
\delta &=  (m^{\rm rec}_{\mu\mu} - m_{\mu\mu})/\sigma_C \\
A_{\rm l, u} & = \left( \frac{n_{\rm l, u}}{|\alpha_{\rm l, u}|} \right)^{n_{\rm l, u}} e^{-|\alpha_{\rm l, u}|^2/2 }\, \\ 
B_{\rm l, u} & = \left(  \frac{n_{\rm l, u}}{|\alpha_{\rm l, u}|} \right) -  |\alpha_{\rm l, u}|
\end{split}
\end{align} 
and is normalised to unity.

\begin{table}[tb]
\caption{
Resolution parameters of the different convolution regions in units of \mevcc. 
The $\alpha_{\rm l}$ and $\alpha_{\rm u}$ parameters are shared between the \jpsi and \psitwos regions.
The parameters without uncertainties are fixed from fits to the simulated events. 
}
\resizebox{\textwidth}{!}{
    \begin{tabular}{ c  c c c c c c c}
      \hline
      Region (\mevcc) & $\sigma_G$ & $\sigma_C$ & $\alpha_{\rm l}$ & $n_{\rm l}$ & $\alpha_{\rm u}$ & $n_{\rm u}$ & $f$  \\
      \hline
	$[\phantom{0}300,1800]$  & 3.53 & 2.98 & $-$1.15 & 20.0 & 1.15 & 20.0 & $0.39$ \\ 
        $[1800,3400]$ & $6.71\pm0.04$ & $5.67\pm0.02$  & $-1.21\pm0.02$ & $9.1\pm1.0$ & $1.21\pm0.02$ & 20.0 & $0.41\pm0.01$ \\ 
        $[3400,4700]$ & $5.63\pm0.04$ & $4.76\pm0.02$ & $-1.21\pm0.02$ & $8.5\pm0.5$ & $1.21\pm0.02$ & $7.3\pm1.2$ & $0.41\pm0.01$ \\ 
      \hline
    \end{tabular}
}\\
\label{tab:resolution_params}
\end{table}

The parameters describing the resolution model for the \jpsi and \psitwos regions ($f$, $\sigma_C$, $\sigma_G$, $n_{\rm l}$, $n_{\rm u}$, $\alpha_{\rm l}$, $\alpha_{\rm u}$) are allowed to vary in the fit to the data. The parameters $\alpha_{\rm l}$, $\alpha_{\rm u}$ and $f$ are shared between the \jpsi and \psitwos regions.
The resolution parameters for the $\phi$ region can not be determined from the data in this way and are instead fixed to their values in the simulation. 
The resulting values of the resolution parameters are summarised in Table~\ref{tab:resolution_params}.
As a cross-check, a second fit to the $m_{\mu\mu}^{\rm rec}$ distribution is performed using the full $m_{\mu\mu}$ dependence of the resolution model in Eq.~\ref{eq:fullmodel} and a numerical implementation of the convolution. 
In this fit to the data, the parameters of the resolution model are taken from simulated \decay{\Bp}{\Kp\mumu} events and fixed up to an overall scaling of the width of the resolution function. 
The two fits to $m_{\mu\mu}^{\rm rec}$ yield compatible results.

\subsection{Efficiency correction}
\label{sec:effic-corr}

\begin{figure}
  \centering
  \includegraphics[width=0.6\textwidth] {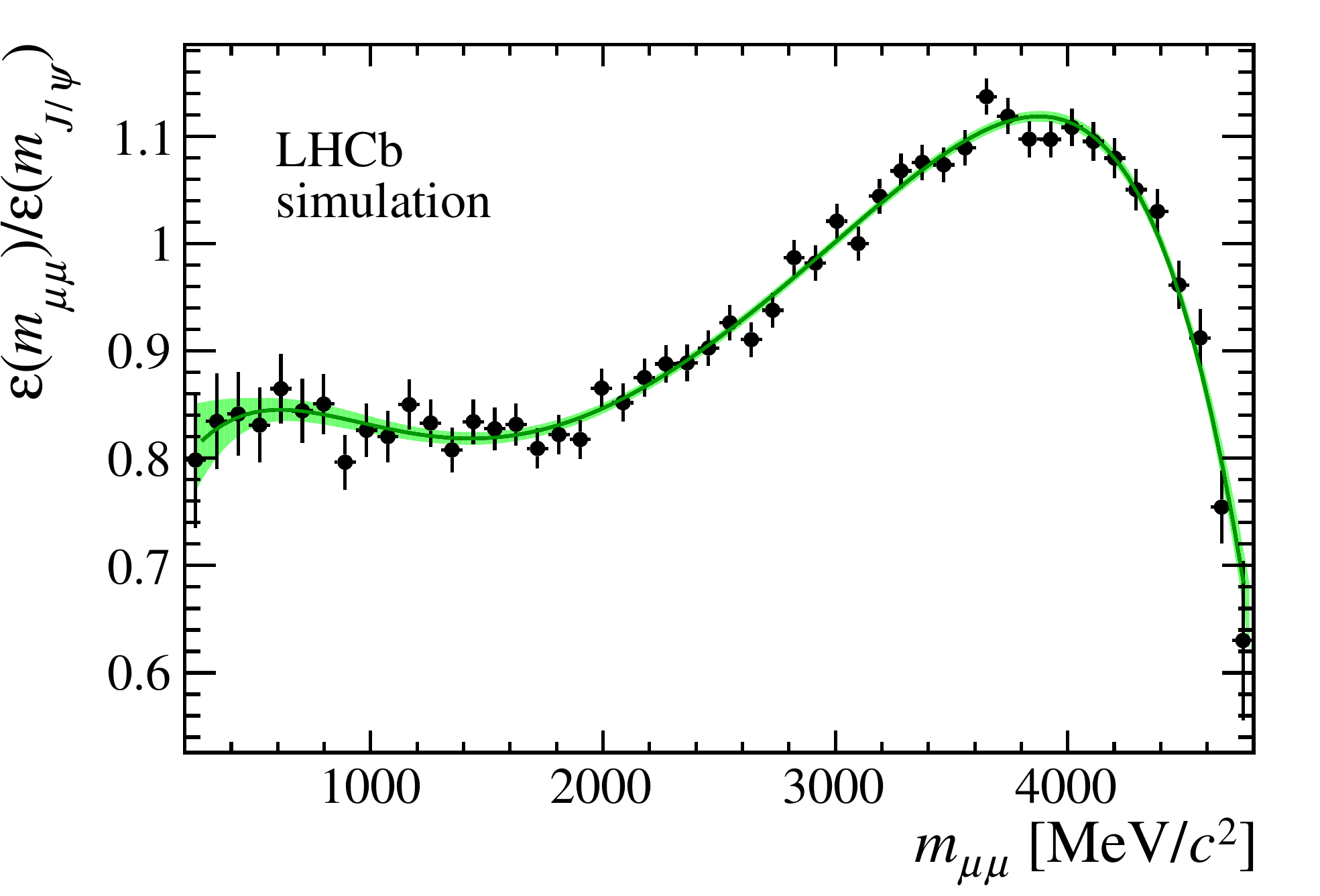}
  \caption{Efficiency to reconstruct, trigger and select
    simulated $\BuKMuMu$ decays as a function of the true dimuon
    mass. The efficiency is normalised to the efficiency at the \jpsi meson mass. The band indicates the  efficiency parameterisation used in this analysis and its statistical uncertainty.}
  \label{fig:eff}
\end{figure}

The measured dimuon mass distribution is biased by the trigger, selection and detector geometry. 
The dominant sources of bias are the geometrical acceptance of the detector,  the impact parameter requirements on the muons and the kaon and the \pt dependence of the trigger. 
Figure~\ref{fig:eff} shows the efficiency to trigger, reconstruct and select candidates as a function of $m_{\mu\mu}$  in a sample of simulated \BuToKmm candidates. 
The rise in efficiency with increasing dimuon mass originates from the requirement that one of the muons has $\pt > 1.48\gevc$ ($\pt > 1.76\gevc)$ in the 2011 (2012) trigger. 
The drop in efficiency at large dimuon mass (small hadronic recoil) originates from the impact parameter requirement on the kaon.
The efficiency is normalised to the efficiency at the \jpsi meson mass and is parameterised 
as a function of $m_{\mu\mu}$ by the sum of Legendre polynomials, $P_i(x)$, up to sixth order,
\begin{align}
\varepsilon(m_{\mu\mu}) = \sum\limits_{i=0}^{6} \varepsilon_{i}   P_i \left(-1 + 2\left(\frac{m_{\mu\mu} - 2m_{\mu}}{m_B - m_K - 2 m_{\mu}}\right) \right).
\end{align}
The values of the parameters $\varepsilon_i$ are fixed from simulated events and are given in Table~\ref{tab:efficiency}. 

\begin{table}
\centering
\caption{ 
Parameters describing the efficiency to trigger, reconstruct and select simulated \decay{\Bp}{\Kp\mumu}  decays as a function of $m_{\mu\mu}$.}
\begin{tabular}{cccccccc}
\hline
& $\varepsilon_0$ & $\varepsilon_1$  & $\varepsilon_2$  & $\varepsilon_3$  & $\varepsilon_4$  & $\varepsilon_5$ &  $\varepsilon_6$ \\ 
\hline
Value & $\phantom{+}0.9262$ &  $\phantom{+}0.1279$ & $-0.0532$  & $-0.1857$   &  $-0.1269$  & $-0.0205$   & $-0.0229$    \\ 
Uncertainty &  $\phantom{+}0.0036$  & $\phantom{+}0.0080$  & $\phantom{+}0.0116$  & $\phantom{+}0.0131$  &  $\phantom{+}0.0155$ &  $\phantom{+}0.0138$  &  $\phantom{+}0.0148$  \\  
\hline
 \\
 \hline
Correlation & $\varepsilon_0$ & $\varepsilon_1$  & $\varepsilon_2$  & $\varepsilon_3$  & $\varepsilon_4$  & $\varepsilon_5$ &  $\varepsilon_6$ \\ 
\hline 
$\varepsilon_0$ & $\phantom{+}1.000$ & $-0.340$  & $\phantom{+}0.605$ & $-0.208$  & $\phantom{+}0.432$ & $-$0.132  & $\phantom{+}0.298$ \\ 
$\varepsilon_1$ &  & $\phantom{+}1.000$ & $-0.345$ & $\phantom{+}0.635$ &  $-0.207$ &  $\phantom{+}0.411$ & $-0.094$  \\ 
$\varepsilon_2$ &  &  & \phantom{+}1.000 & $-0.352$ &  $\phantom{+}0.684$ & $-0.224$ &  $\phantom{+}0.455$ \\ 
$\varepsilon_3$ &  &  &  & $\phantom{+}1.000$ & $-0.344$ & $\phantom{+}0.608$ & $-0.154$ \\ 
$\varepsilon_4$ &  &  &  &  & $\phantom{+}1.000$ & $-0.344$ &  $\phantom{+}0.619$ \\ 
$\varepsilon_5$ &  &  &  &  &  & $\phantom{+}1.000$ & $-0.259$ \\ 
$\varepsilon_6$ &  &  &  &  &  &  & $\phantom{+}1.000$ \\ 
\hline
\end{tabular} 
\label{tab:efficiency} 
\end{table}

\subsection{Background model}

The reconstructed dimuon mass distribution of the combinatorial background candidates is taken from the $m_{K\mu\mu}$ upper mass sideband, $5620 < m_{K\mu\mu} < 5700\mevcc$. 
When evaluating $m^{\rm rec}_{\mu\mu}$,  $m_{K\mu\mu}$  is constrained to the centre of the sideband rather than to the known \Bp mass. 
Combinatorial background comprising a genuine \jpsi or \psitwos meson is described by the sum of two Gaussian functions. 
After applying the mass constraint, the means of the Gaussians do not correspond exactly to the known \jpsi and \psitwos masses. 
Combinatorial background comprising a dimuon pair that does not originate from a \jpsi or \psitwos meson is described by an ARGUS function~\cite{Albrecht:1994tb}.
The lineshape of the background from \decay{\Bp}{\pip\mumu} decays, where the pion is mistakenly identified as a kaon, is taken from simulated events.

\section{Results}
\label{sec:results}

\begin{figure}[tb]
\centering
\includegraphics[width=0.49\textwidth] {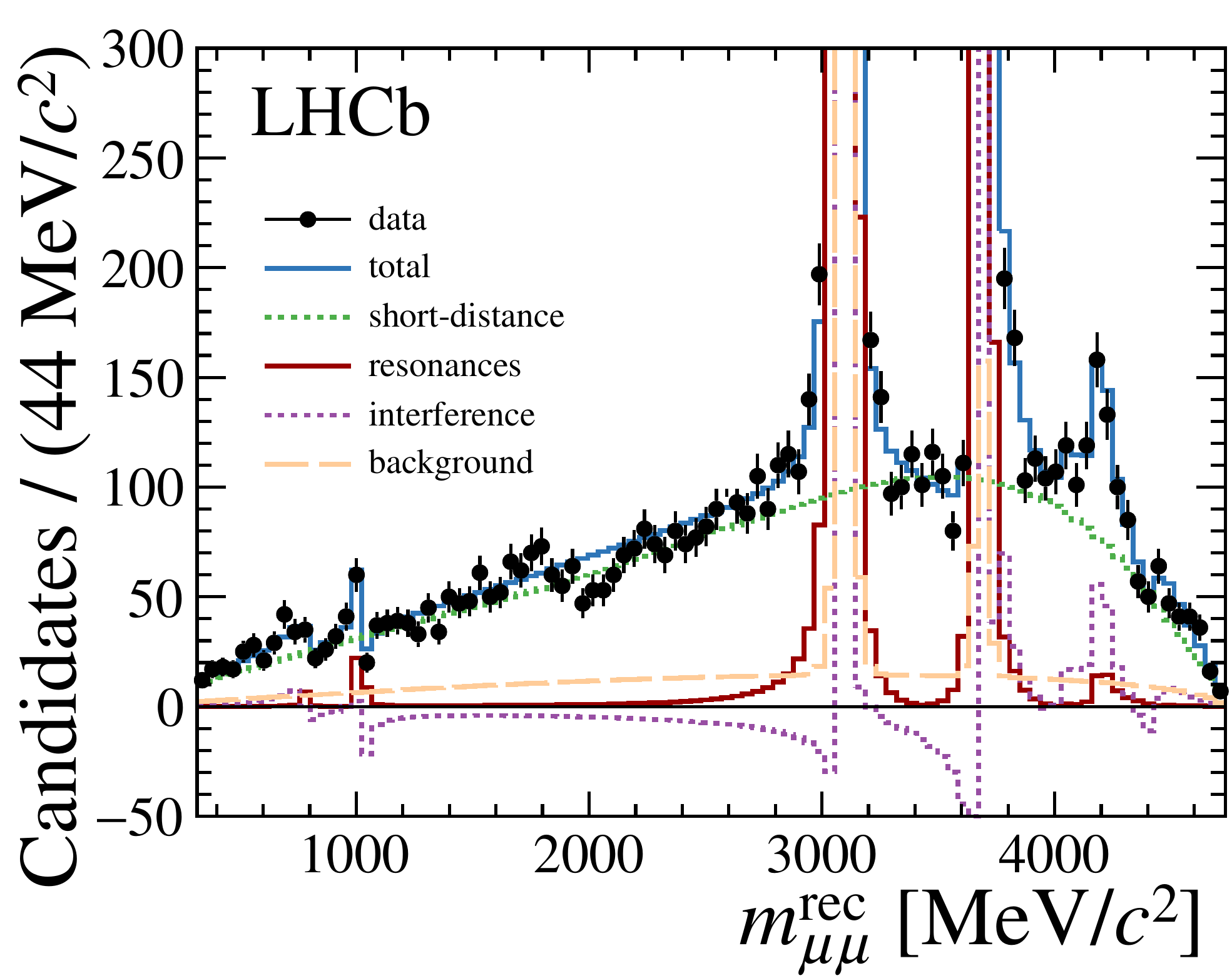}
\hspace*{-0.35cm}
\includegraphics[width=0.49\textwidth] {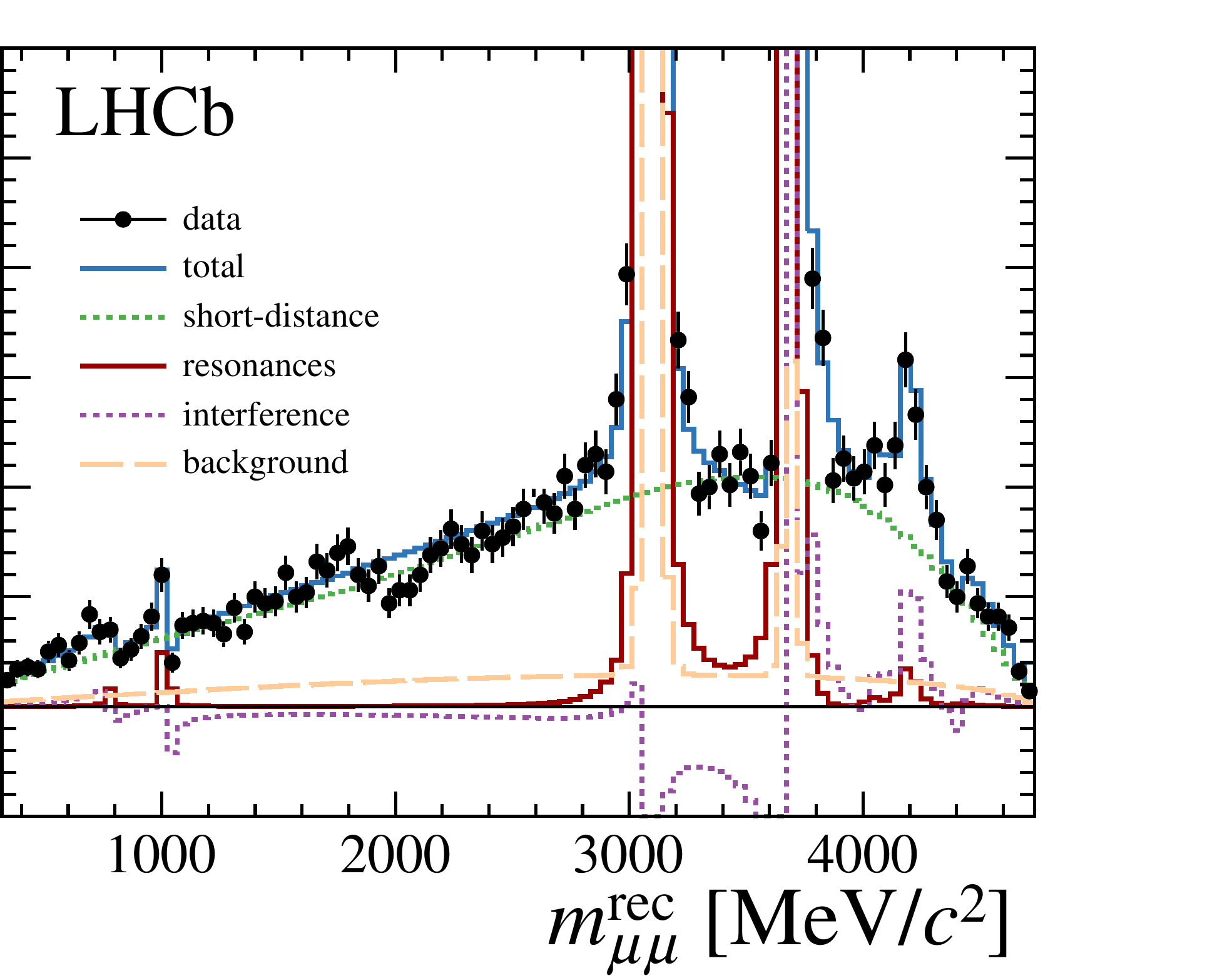}
\includegraphics[width=0.49\textwidth] {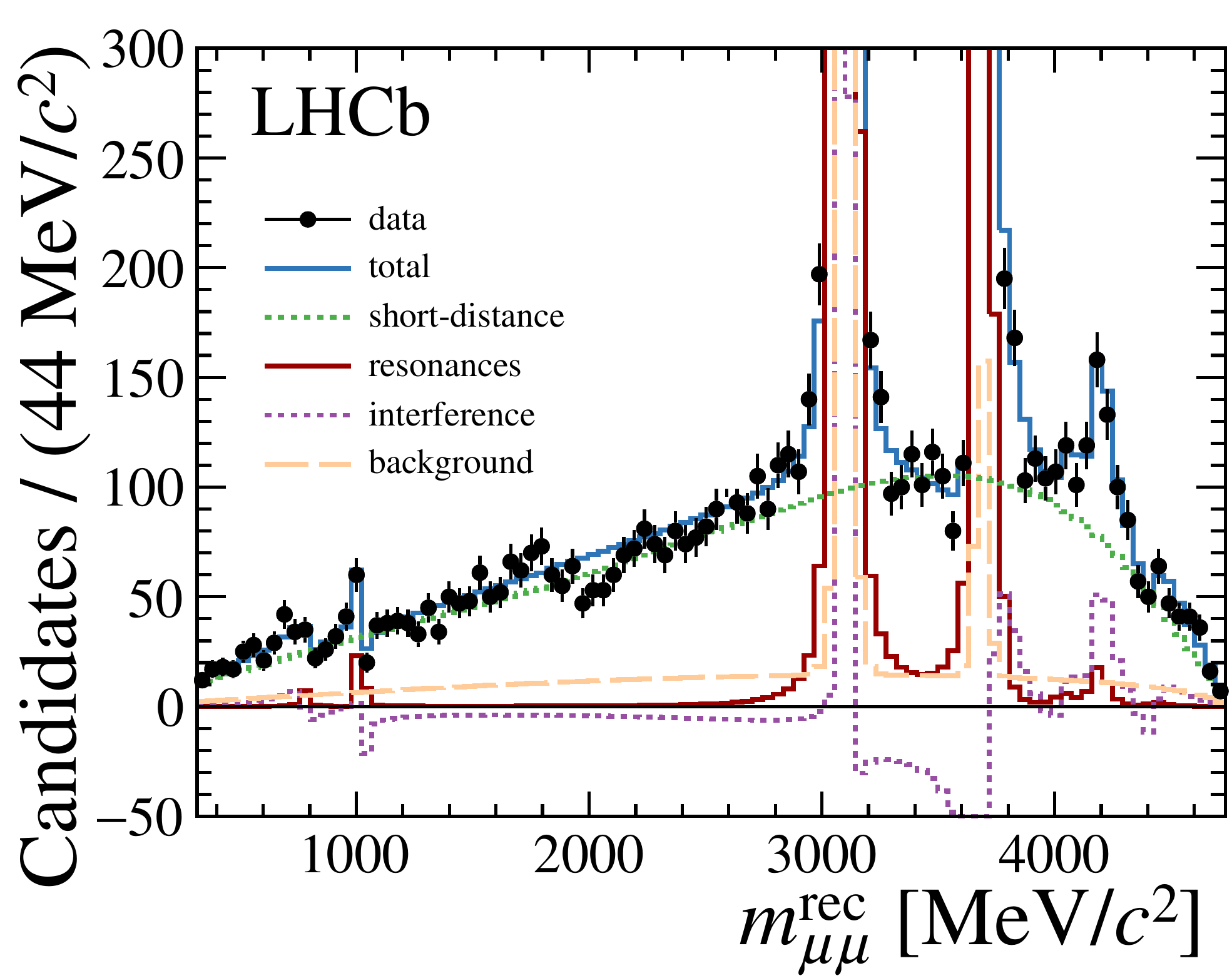}
\hspace*{-0.35cm}
\includegraphics[width=0.49\textwidth] {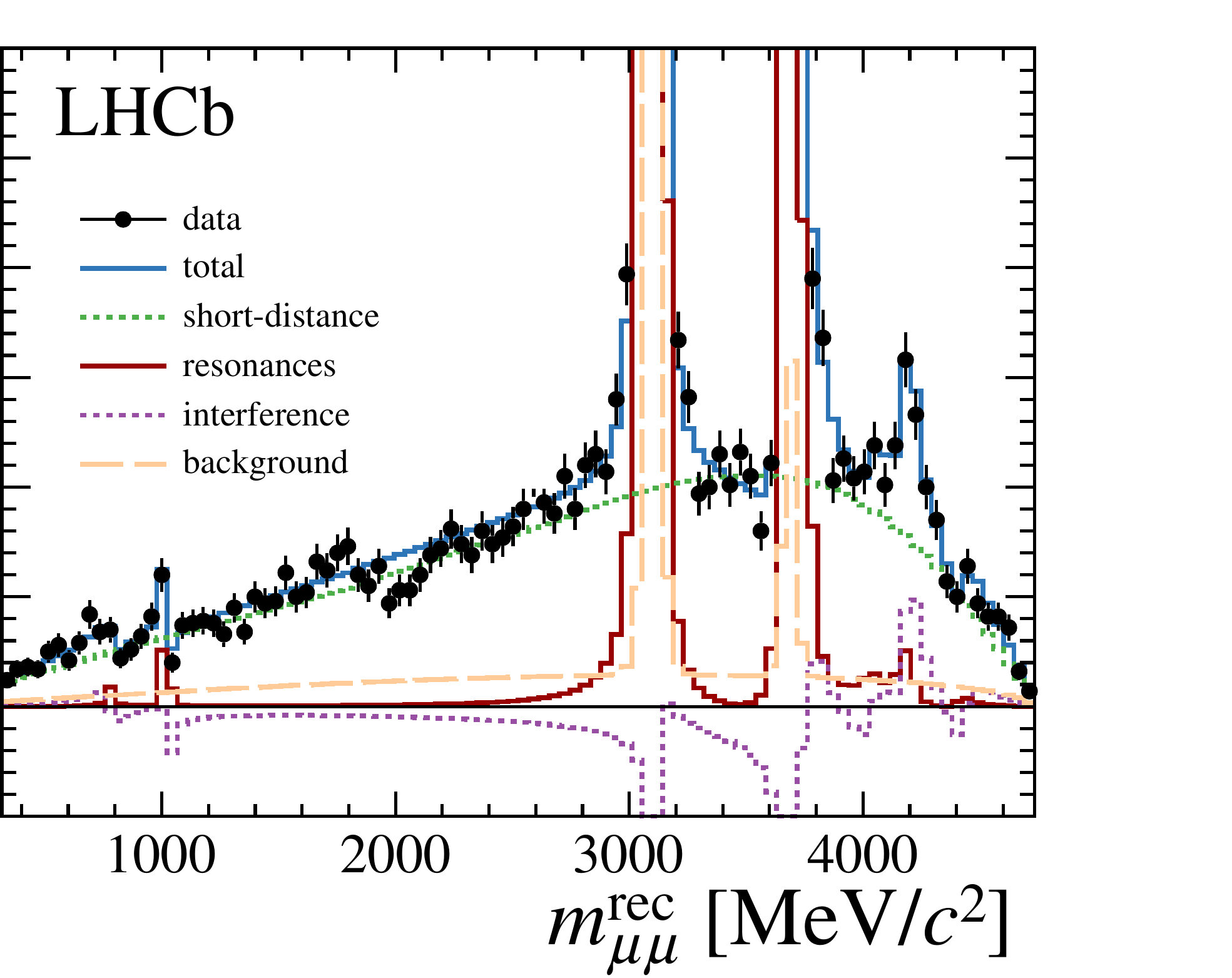}
\caption{
Fits to the dimuon mass distribution for the four different phase combinations that describe the data equally well. The plots show cases where the \jpsi and \psitwos phases are both negative (top left); 
the \jpsi phase is positive and the \psitwos phase is negative (top right); 
the \jpsi phase is negative and the \psitwos phase is positive (bottom left); 
and both phases are positive (bottom right). The component labelled interference refers to the interference between the short- and long-distance contributions to the decay. The \chisq value of the four solutions is almost identical, with a value of 110 for 78 degrees of freedom.
}
\label{fig:finalfit}
\end{figure}

The dimuon mass distributions and the projections of the fit to the data are shown in Fig.~\ref{fig:finalfit}.
Four solutions are obtained with almost equal likelihood values, which 
correspond to ambiguities in the signs of the \jpsi and \psitwos phases.
The values of the phases and branching fractions of the vector meson resonances are listed in Table~\ref{tab:result_neg_neg}.
The posterior values for the $f_+$ form factor are reported in Table~\ref{tab:results:ff:posterior}.
A \chisq test between the data and the model, with the binning scheme used in Fig.~\ref{fig:finalfit}, results in a \chisq of 
110 with 78 degrees of freedom. The largest disagreements between the data and the model are localised in the 
$m_{\mu\mu}$ region close to the \jpsi pole mass and around 1.8\gevcc. The latter is discussed in Sec.~\ref{sec:systematics}.

\begin{table}[t]
\centering
\caption{Branching fractions and phases for each resonance in the fit for the four solutions of the \jpsi and \psitwos phases. 
Both statistical and systematic contributions are included in the uncertainties. 
There is a common systematic uncertainty of 4.5\%, dominated by the uncertainty on the $\Bp\to\jpsi\Kp$ branching fraction, which provides the normalisation for all measurements.}
{\small 
    \begin{tabular}{ c | c  c | c c}
      \hline
       & \multicolumn{2}{c|}{\jpsi negative/\psitwos negative} & \multicolumn{2}{c}{\jpsi negative/\psitwos positive}  \\
      Resonance & Phase [rad] & Branching fraction & Phase [rad] & Branching fraction \\
     \hline \\ [-2.5ex]
$\rho(770)$ & $-0.35 \pm 0.54$ & $(1.71 \pm 0.25) \times 10^{-10}$ & $-0.30 \pm 0.54$ & $(1.71 \pm 0.25)\times 10^{-10}$ \\
$\omega(782)$ & $\phantom{+}0.26 \pm 0.39$ & $(4.93  \pm 0.59) \times 10^{-10}$ & $\phantom{+}0.30 \pm 0.38$ & $(4.93 \pm 0.58)\times 10^{-10}$ \\
$\phi(1020)$ & $\phantom{+}0.47 \pm 0.39$ & $ (2.53  \pm 0.26) \times 10^{-9\phantom{0}}$ & $\phantom{+}0.51 \pm 0.37$ & $(2.53 \pm 0.26)\times 10^{-9\phantom{0}}$ \\
$J/\psi$ & $-1.66 \pm 0.05$ & -- & $-1.50 \pm 0.05$ & -- \\
$\psi(2S)$ & $-1.93 \pm 0.10$ & $(4.64 \pm 0.20)\times 10^{-6\phantom{0}}$ & $\phantom{+}2.08 \pm 0.11$ & $(4.69 \pm 0.20)\times 10^{-6\phantom{0}}$ \\
$\psi(3770)$ & $-2.13 \pm 0.42$ & $(1.38 \pm 0.54) \times 10^{-9\phantom{0}}$ & $-2.89 \pm 0.19$ & $(1.67 \pm 0.61)\times 10^{-9\phantom{0}}$ \\
$\psi(4040)$ & $-2.52 \pm 0.66$ & $(4.17 \pm 2.72)\times 10^{-10}$ & $-2.69 \pm 0.52$ & $(4.25 \pm 2.83)\times 10^{-10}$ \\
$\psi(4160)$ & $-1.90 \pm 0.64$ & $(2.61 \pm 0.84) \times 10^{-9\phantom{0}}$ & $-2.13 \pm 0.33$ & $(2.67 \pm 0.85)\times 10^{-9\phantom{0}}$ \\
$\psi(4415)$ & $-2.52 \pm 0.36$ & $(6.04 \pm 3.93) \times 10^{-10}$ & $-2.43 \pm 0.43$ & $(7.10 \pm 4.48) \times 10^{-10}$ \\
      \hline
  & \multicolumn{2}{c|}{\jpsi positive/\psitwos negative} & \multicolumn{2}{c}{\jpsi positive/\psitwos positive}  \\
      Resonance & Phase [rad] & Branching fraction & Phase [rad] & Branching fraction \\
      \hline \\ [-2.5ex]
$\rho(770)$ & $-0.26 \pm 0.54$ & $(1.71 \pm 0.25)\times 10^{-10}$ & $-0.22 \pm 0.54$ & $(1.71 \pm 0.25) \times 10^{-10}$ \\
$\omega(782)$ & $\phantom{+}0.35 \pm 0.39$ & $(4.93 \pm 0.58)\times 10^{-10}$ & $\phantom{+}0.38 \pm 0.38$ & $(4.93  \pm 0.58) \times 10^{-10}$ \\
$\phi(1020)$ & $\phantom{+}0.58 \pm 0.38$ & $(2.53 \pm 0.26)\times 10^{-9\phantom{0}}$ & $\phantom{+}0.62 \pm 0.37$ & $(2.52 \pm 0.26)\times 10^{-9\phantom{0}}$ \\
$J/\psi$ & $\phantom{+}1.47 \pm 0.05$ & -- & $\phantom{+}1.63 \pm 0.05$ & -- \\
$\psi(2S)$ & $-2.21 \pm 0.11$ & $(4.63 \pm 0.20)\times 10^{-6\phantom{0}}$ & $\phantom{+}1.80 \pm 0.10$ & $(4.68 \pm 0.20) \times 10^{-6\phantom{0}}$ \\
$\psi(3770)$ & $-2.40 \pm 0.39$ & $(1.39 \pm  0.54)\times 10^{-9\phantom{0}}$ & $-2.95 \pm 0.14$ & $(1.68 \pm 0.61) \times 10^{-9\phantom{0}}$ \\
$\psi(4040)$ & $-2.64 \pm 0.50$ & $(4.05 \pm 2.76)\times 10^{-10}$ & $-2.75 \pm 0.48$ & $(4.30 \pm 2.86)  \times 10^{-10}$ \\
$\psi(4160)$ & $-2.11 \pm 0.38$ & $(2.62 \pm 0.82)\times 10^{-9\phantom{0}}$ & $-2.28 \pm 0.24$ & $(2.68 \pm 0.81) \times 10^{-9\phantom{0}}$ \\
$\psi(4415)$ & $-2.42 \pm 0.46$ & $(6.13 \pm 3.98)\times 10^{-10}$ & $-2.31 \pm 0.48$ & $(7.12 \pm 4.94) \times 10^{-10}$ \\
\hline
\end{tabular}
}
\label{tab:result_neg_neg}
\end{table}

\begin{table}[b]
\centering
\caption{Coefficients of the form factor $f_{+}(\qsq)$ as introduced in Eq.~\ref{eq:FF} with both prior (from Ref.~\cite{Bailey:2015dka}) and posterior values shown.}
    \begin{tabular}{ c  c  c }
      \hline
      Coefficient & Ref.~\cite{Bailey:2015dka} & Fit result \\
      \hline
        $b^{+}_{0}$ & $\phantom{+}0.466\pm0.014$ & $\phantom{+}0.465\pm0.013$ \\ 
        $b^{+}_{1}$ & $-0.89\pm0.13$ & $-0.81\pm0.05$ \\ 
        $b^{+}_{2}$ & $-0.21\pm0.55$ & $\phantom{+}0.03\pm0.32$ \\ 
      \hline
    \end{tabular}
\label{tab:results:ff:posterior}
\end{table}

The branching fraction of  the short-distance component of the \BuKMuMu decay can be calculated by integrating Eq.~\ref{eq:dGamma} after setting the amplitudes of the resonances to zero. 
This gives
\begin{displaymath}
\mathcal{B}(\BuKMuMu) = (4.37\pm0.15\stat\pm0.23\syst) \times 10^{-7}\; ,
\end{displaymath}
where the statistical uncertainty includes the  uncertainty on the form-factor predictions.
The systematic uncertainty on the branching fraction is discussed in Sec.~\ref{sec:systematics}.
This measurement is compatible with the branching fraction reported in Ref.~\cite{LHCb-PAPER-2014-006}. 
The two results are based on the same data and therefore should not be used together in global fits. The branching fraction reported in Ref.~\cite{LHCb-PAPER-2014-006} is based on a binned measurement in \qsq regions away from the narrow resonances ($\phi$, \jpsi and \psitwos) and then  extrapolated to the full \qsq range. The contribution from the broad resonances was thus included in that result.

\begin{figure}[tb]
\centering
\includegraphics[width=0.65\linewidth]{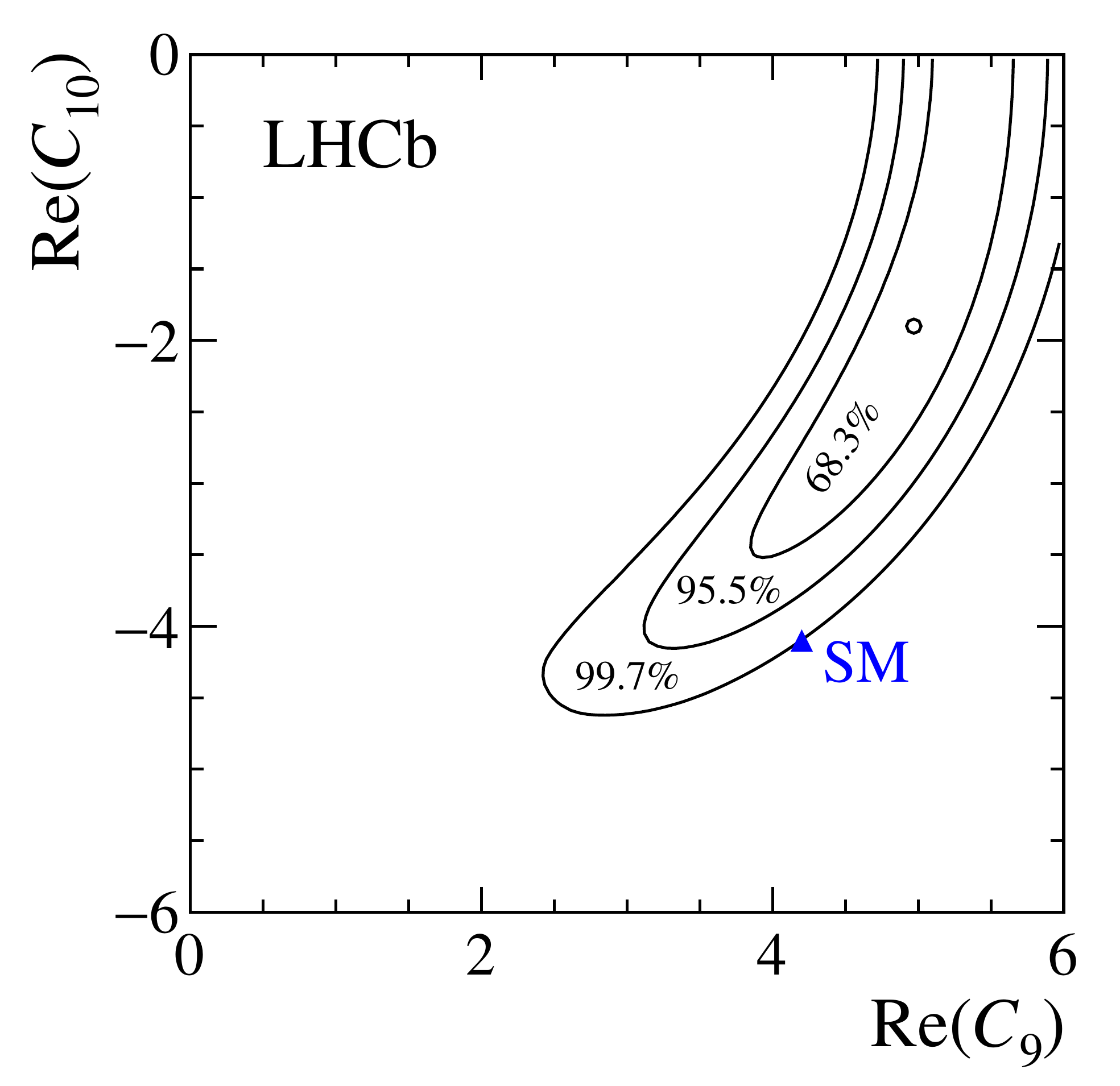}
\caption{
Two-dimensional likelihood profile for the Wilson coefficients $\mathcal{C}_{9}$ and  $\mathcal{C}_{10}$. 
The SM point is indicated by the blue marker. 
The intervals correspond to \chisq probabilities with two degrees of freedom.
\label{fig:twodprofile}
}
\end{figure}

A two-dimensional likelihood profile of $\mathcal{C}_{9}$ and $\mathcal{C}_{10}$ is also obtained as shown in Fig.~\ref{fig:twodprofile}. 
The intervals correspond to \chisq probabilities assuming two degrees of freedom. 
Only the quadrant with $\mathcal{C}_{9}$ and $\mathcal{C}_{10}$ values around the SM prediction is shown. 
The other quadrants can be obtained by mirroring in the axes. 
The branching fraction of the short-distance component provides a good constraint on the sum of 
$|\mathcal{C}_{9}|^{2}$ and $|\mathcal{C}_{10}|^{2}$ (see Eq.~\ref{eq:dGamma}). 
This gives rise to the annular shape in the likelihood profile in Fig.~\ref{fig:twodprofile}. 
In addition, there is a modest ability for the fit to differentiate between $\mathcal{C}_{9}$ and $\mathcal{C}_{10}$ through the interference of the  $\mathcal{C}_{9}$ component with the resonances.
The visible interference pattern excludes very small values of $|\mathcal{C}_{9}|$.
Overall, the correlation between $\mathcal{C}_{9}$ and $\mathcal{C}_{10}$ is approximately 90\%. 
The best-fit point for the Wilson coefficients (in a given quadrant of the $\mathcal{C}_{9}$ and $\mathcal{C}_{10}$ plane) and the corresponding \decay{\Bp}{\Kp\mumu} branching fraction are the same for the four combinations of the \jpsi and \psitwos phases. 
Including statistical and systematic uncertainties, the fit  results deviate from the SM prediction at the level of 3.0 standard deviations. 
The uncertainty is dominated by the precision of the form factors.
The best-fit point prefers a value of $|\mathcal{C}_{10}|$ that is smaller than $|\mathcal{C}_{10}^{\rm SM}|$ and a value of $|\mathcal{C}_{9}|$ that is larger than $|\mathcal{C}_{9}^{\rm SM}|$.
However, if $\mathcal{C}_{10}$ is fixed to its SM value, the fit prefers  $|\mathcal{C}_{9}| <  |\mathcal{C}_{9}^{\rm SM}|$. 
This is consistent with the results of global fits to \decay{b}{s\ell^+\ell^-} processes. 
Given the model assumptions in this paper, the interference with the \jpsi meson is not able to explain the low value of the branching fraction of the \decay{\Bp}{\Kp\mumu} decay while keeping the values of \Cnine and $\mathcal{C}_{10}$ at their SM predictions.

\section{Systematic uncertainties}
\label{sec:systematics}

Sources of systematic uncertainty are considered separately for the phase
 and branching fraction measurements. 
 In both cases, the largest systematic uncertainties are accounted for in the statistical uncertainty as they are included as nuisance parameters in the fit. 
 For smaller sources of uncertainty, the fit is repeated with variations of the inputs and the difference is assigned as a systematic uncertainty.
 A summary of the remaining systematic uncertainties can be found in Table~\ref{tab:systsummary}.
  
  \begin{table}[b]
  \caption{Summary of systematic uncertainties. 
  The branching fraction refers to the short-distance SM contribution. A dash indicates that the uncertainty is negligible.}
\centering
    \begin{tabular}{ l  c  c  c c}
      \hline
      Source & \jpsi phase & \psitwos phase & Branching fraction & $\mathcal{C}_{9,10}$ \\
      \hline
        Broad components & 20\mrad & 10\mrad & 1.0\% & 0.05 \\
        Background model & 10\mrad & 10\mrad & 1.0\% & 0.05 \\ 
        Efficiency model & \phantom{0}3\mrad & 10\mrad & 1.0\% & 0.05 \\ 
        $\BF(\BuJpsiK)$ & --- & --- & 4.2\% & 0.19 \\
      \hline
    \end{tabular}
\label{tab:systsummary}
\end{table}
  
The parameters governing the behaviour of 
  the tails of the resolution function are particularly correlated with the phases. 
  The systematic uncertainty on the resolution model is included in the statistical 
  uncertainty by allowing the resolution parameter values to vary in the fit. 
  If the tail parameters are fixed to their central values, the statistical uncertainties 
  on the phase measurements 
  decrease by approximately 20\%.
   The choice of parameterisation for the 
   resolution model is validated using a large sample of simulated events and no additional uncertainty is assigned for the choice of model. 
   For the branching fraction measurement, the uncertainty arising from the resolution model
   is negligible compared to other sources of systematic uncertainty.
   
    Similarly to the resolution model, the systematic uncertainty associated with the knowledge
     of the $f_{+}(\qsq)$ form factor is included in the statistical uncertainty. 
     If the form-factor parameters are fixed to their best-fit values, the statistical uncertainties on the phases decrease by 4\% (1\%)
  for the \jpsi (\psitwos) measurements. 
    For the branching fraction, the uncertainty is 2\%, which is of similar size as the statistical uncertainty. 
  
  At around $m_{\mu\mu} =1.8$\gevcc  there is a small discrepancy between
  the data and the model (see Fig.~\ref{fig:finalfit}). This is interpreted as a possible contribution from excited 
  $\rho$, $\omega$ or $\phi$ resonances. Given the limited knowledge of the
  masses and widths of the states in this region, these broad states are neglected in the
   nominal fit. They are, however, visible in $\epem\to \rm hadrons$ vacuum polarisation data~\cite{Olive:2016xmw}.
    To test the effect of such states on the phases of the \jpsi and \psitwos mesons, an additional 
   relativistic Breit--Wigner amplitude is included with a width and mass that are allowed to vary in the fit. 
   The inclusion of this Breit--Wigner amplitude marginally improves the fit quality around $m_{\mu\mu} =1.8$\gevcc and changes 
   the \jpsi (\psitwos) phase by 40\% (20\%) of its statistical uncertainty, which is added as a systematic effect. 
   The magnitude of the amplitude is not statistically significant and its mean and width do not correspond to a known state.
   The phases of the other resonances in the fit have larger statistical uncertainties and the inclusion of this additional amplitude has a negligible effect on their fit values.
   Given that the contribution of this amplitude is small compared 
   to the short-distance component, its effect on the branching fraction is only around 1\%.
   
   Other, smaller systematic uncertainties include modelling of the combinatorial 
   background, calculation of the efficiency as a function of \qsq and the uncertainty on the 
   \BuJpsiK branching fraction. The latter affects the branching fraction measurement 
   and is obtained from Ref.~\cite{Jung:2015yma}, which
   results in a $4\%$ uncertainty.

\section{Conclusions}
\label{sec:conclusions}

This paper presents the first measurement of the phase difference between the short- and long-distance contributions to the $\BuToKmm$ decay.
The measurement is performed using a binned maximum likelihood fit to the dimuon mass distribution of the decays. 
The long-distance contributions are modelled as the sum of relativistic Breit--Wigner amplitudes representing different vector meson resonances decaying to muon pairs, each with their own magnitude and phase. The short-distance contribution is expressed in terms of an effective field theory description of the decay with the Wilson coefficients $\mathcal{C}_9$ and $\mathcal{C}_{10}$, which are taken to be real. 
These are left free in the fit and all other components set to their corresponding SM values. 
The $B\to K$ hadronic form factors are constrained in the fit to the predictions from Ref.~\cite{Bailey:2015dka}.

The fit results in four approximately degenerate solutions corresponding to ambiguities in the signs of the \jpsi and \psitwos phases. The values of the \jpsi phases are compatible with $\pm \tfrac{\pi}{2}$, which means that the interference with the short-distance component in dimuon mass regions far from their pole masses is small. The negative solution of the \jpsi phase agrees qualitatively with the prediction of Ref.~\cite{Khodjamirian:2012rm}, where long-distance contributions are calculated at negative \qsq and extrapolated to the \qsq region below the \jpsi pole-mass using a hadronic dispersion relation. The fit model, which includes the conventional $J^{PC}=1^{--}$ $c\bar{c}$ resonances, is found to describe the data well, with no significant evidence for the decays \decay{\Bp}{\psi(4040)\Kp} or \decay{\Bp}{\psi(4415)\Kp}. The values of the $\psi(3770)$ and $\psi(4160)$ phases are compatible with those reported in Ref.~\cite{Lyon:2014hpa}.

The measurement of the Wilson coefficients prefers a value of $|\mathcal{C}_{10}| < |\mathcal{C}_{10}^{\rm SM}|$ and a value of $|\mathcal{C}_{9}| > |\mathcal{C}_{9}^{\rm SM}|$. If the value of $\mathcal{C}_{10}$ is set to that of $\mathcal{C}_{10}^{\rm SM}$, the measurement favours the region $|\mathcal{C}_{9}| < |\mathcal{C}_{9}^{\rm SM}|$. These results are similar to those reported previously in global analyses. The interference between the short- and long-distance contributions in the regions around the $\rho$, $\omega$ and the $\phi$, and in the region $\qsq > m_{\psitwos}^2$, results in the exclusion of the hypothesis that $\mathcal{C}_{9} = 0$ at more than 5 standard deviations. The dominant uncertainty on the measurements of $\mathcal{C}_{9}$ and $\mathcal{C}_{10}$ arises from the knowledge of the $B\to K$ hadronic form factors. The current data set allows the uncertainties on these hadronic parameters to be reduced. Improved inputs on the form factors from lattice QCD calculations and the larger data set that will be available at the end of the LHC Run\,2 are needed to further improve the measurement of the Wilson coefficients.

A similar strategy to the one applied in this paper can be extended to other $b\to s\ell^+\ell^-$ decay processes to understand the influence of hadronic resonances on global fits for \Cnine and $\mathcal{C}_{10}$. However, the situation is more complicated in decays where the strange hadron is not a pseudoscalar meson as the amplitudes corresponding to different helicity states of the hadron can have different relative phases.

Finally, a measurement of the branching fraction of the short-distance component of $\BuToKmm$ decays is also reported and is found to be
\begin{displaymath}
\mathcal{B}(\BuKMuMu) = (4.37\pm0.15\stat\pm0.23\syst) \times 10^{-7}\; ,
\end{displaymath}
where the first uncertainty is statistical and second is systematic. In contrast to previous analyses, 
the measurement is performed across the full \qsq region accounting for the interference with the 
long-distance contributions and without any veto of resonance-dominated regions of the phase space. 
The value of the branching fraction is found 
to be compatible with previous measurements~\cite{LHCb-PAPER-2014-006}, but smaller than the SM prediction~\cite{Bailey:2015dka}.

\section*{Acknowledgements}
 
\noindent We express our gratitude to our colleagues in the CERN
accelerator departments for the excellent performance of the LHC. We
thank the technical and administrative staff at the LHCb
institutes. We acknowledge support from CERN and from the national
agencies: CAPES, CNPq, FAPERJ and FINEP (Brazil); NSFC (China);
CNRS/IN2P3 (France); BMBF, DFG and MPG (Germany); INFN (Italy);
FOM and NWO (The Netherlands); MNiSW and NCN (Poland); MEN/IFA (Romania);
MinES and FASO (Russia); MinECo (Spain); SNSF and SER (Switzerland);
NASU (Ukraine); STFC (United Kingdom); NSF (USA).
We acknowledge the computing resources that are provided by CERN, IN2P3 (France), KIT and DESY (Germany), INFN (Italy), SURF (The Netherlands), PIC (Spain), GridPP (United Kingdom), RRCKI and Yandex LLC (Russia), CSCS (Switzerland), IFIN-HH (Romania), CBPF (Brazil), PL-GRID (Poland) and OSC (USA). We are indebted to the communities behind the multiple open
source software packages on which we depend.
Individual groups or members have received support from AvH Foundation (Germany),
EPLANET, Marie Sk\l{}odowska-Curie Actions and ERC (European Union),
Conseil G\'{e}n\'{e}ral de Haute-Savoie, Labex ENIGMASS and OCEVU,
R\'{e}gion Auvergne (France), RFBR and Yandex LLC (Russia), GVA, XuntaGal and GENCAT (Spain), Herchel Smith Fund, The Royal Society, Royal Commission for the Exhibition of 1851 and the Leverhulme Trust (United Kingdom).



\addcontentsline{toc}{section}{References}
\setboolean{inbibliography}{true}
\bibliographystyle{LHCb}

\begin{mcitethebibliography}{10}
\mciteSetBstSublistMode{n}
\mciteSetBstMaxWidthForm{subitem}{\alph{mcitesubitemcount})}
\mciteSetBstSublistLabelBeginEnd{\mcitemaxwidthsubitemform\space}
{\relax}{\relax}

\bibitem{Descotes-Genon:2013wba}
S.~Descotes-Genon, J.~Matias, and J.~Virto,
  \ifthenelse{\boolean{articletitles}}{\emph{{Understanding the $B \to
  K^*\mu^+\mu^-$ anomaly}},
  }{}\href{http://dx.doi.org/10.1103/PhysRevD.88.074002}{Phys.\ Rev.\
  \textbf{D88} (2013) 074002},
  \href{http://arxiv.org/abs/1307.5683}{{\normalfont\ttfamily
  arXiv:1307.5683}}\relax
\mciteBstWouldAddEndPuncttrue
\mciteSetBstMidEndSepPunct{\mcitedefaultmidpunct}
{\mcitedefaultendpunct}{\mcitedefaultseppunct}\relax
\EndOfBibitem
\bibitem{Altmannshofer:2013foa}
W.~Altmannshofer and D.~M. Straub,
  \ifthenelse{\boolean{articletitles}}{\emph{{New physics in $B \to
  K^*\mu\mu$?}},
  }{}\href{http://dx.doi.org/10.1140/epjc/s10052-013-2646-9}{Eur.\ Phys.\ J.\
  \textbf{C73} (2013) 2646},
  \href{http://arxiv.org/abs/1308.1501}{{\normalfont\ttfamily
  arXiv:1308.1501}}\relax
\mciteBstWouldAddEndPuncttrue
\mciteSetBstMidEndSepPunct{\mcitedefaultmidpunct}
{\mcitedefaultendpunct}{\mcitedefaultseppunct}\relax
\EndOfBibitem
\bibitem{Altmannshofer:2014cfa}
W.~Altmannshofer, S.~Gori, M.~Pospelov, and I.~Yavin,
  \ifthenelse{\boolean{articletitles}}{\emph{{Quark flavor transitions in
  $L_\mu - L_\tau$ models}},
  }{}\href{http://dx.doi.org/10.1103/PhysRevD.89.095033}{Phys.\ Rev.\
  \textbf{D89} (2014) 095033},
  \href{http://arxiv.org/abs/1403.1269}{{\normalfont\ttfamily
  arXiv:1403.1269}}\relax
\mciteBstWouldAddEndPuncttrue
\mciteSetBstMidEndSepPunct{\mcitedefaultmidpunct}
{\mcitedefaultendpunct}{\mcitedefaultseppunct}\relax
\EndOfBibitem
\bibitem{Mahmoudi:2014mja}
F.~Mahmoudi, S.~Neshatpour, and J.~Virto,
  \ifthenelse{\boolean{articletitles}}{\emph{{$B \to K^{*} \mu^{+} \mu^{-}$
  optimised observables in the MSSM}},
  }{}\href{http://dx.doi.org/10.1140/epjc/s10052-014-2927-y}{Eur.\ Phys.\ J.\
  \textbf{C74} (2014) 2927},
  \href{http://arxiv.org/abs/1401.2145}{{\normalfont\ttfamily
  arXiv:1401.2145}}\relax
\mciteBstWouldAddEndPuncttrue
\mciteSetBstMidEndSepPunct{\mcitedefaultmidpunct}
{\mcitedefaultendpunct}{\mcitedefaultseppunct}\relax
\EndOfBibitem
\bibitem{Crivellin:2015mga}
A.~Crivellin, G.~D'Ambrosio, and J.~Heeck,
  \ifthenelse{\boolean{articletitles}}{\emph{{Explaining $h\to\mu^\pm\tau^\mp$,
  $B\to K^* \mu^+\mu^-$ and $B\to K \mu^+\mu^-/B\to K e^+e^-$ in a
  two-Higgs-doublet model with gauged $L_\mu-L_\tau$}},
  }{}\href{http://dx.doi.org/10.1103/PhysRevLett.114.151801}{Phys.\ Rev.\
  Lett.\  \textbf{114} (2015) 151801},
  \href{http://arxiv.org/abs/1501.00993}{{\normalfont\ttfamily
  arXiv:1501.00993}}\relax
\mciteBstWouldAddEndPuncttrue
\mciteSetBstMidEndSepPunct{\mcitedefaultmidpunct}
{\mcitedefaultendpunct}{\mcitedefaultseppunct}\relax
\EndOfBibitem
\bibitem{Descotes-Genon:2015uva}
S.~Descotes-Genon, L.~Hofer, J.~Matias, and J.~Virto,
  \ifthenelse{\boolean{articletitles}}{\emph{{Global analysis of $b\to
  s\ell\ell$ anomalies}},
  }{}\href{http://dx.doi.org/10.1007/JHEP06(2016)092}{JHEP \textbf{06} (2016)
  092}, \href{http://arxiv.org/abs/1510.04239}{{\normalfont\ttfamily
  arXiv:1510.04239}}\relax
\mciteBstWouldAddEndPuncttrue
\mciteSetBstMidEndSepPunct{\mcitedefaultmidpunct}
{\mcitedefaultendpunct}{\mcitedefaultseppunct}\relax
\EndOfBibitem
\bibitem{Hurth:2016fbr}
T.~Hurth, F.~Mahmoudi, and S.~Neshatpour,
  \ifthenelse{\boolean{articletitles}}{\emph{{On the anomalies in the latest
  LHCb data}},
  }{}\href{http://dx.doi.org/10.1016/j.nuclphysb.2016.05.022}{Nucl.\ Phys.\
  \textbf{B909} (2016) 737},
  \href{http://arxiv.org/abs/1603.00865}{{\normalfont\ttfamily
  arXiv:1603.00865}}\relax
\mciteBstWouldAddEndPuncttrue
\mciteSetBstMidEndSepPunct{\mcitedefaultmidpunct}
{\mcitedefaultendpunct}{\mcitedefaultseppunct}\relax
\EndOfBibitem
\bibitem{Jager:2012uw}
S.~{J\"ager} and J.~Martin~Camalich,
  \ifthenelse{\boolean{articletitles}}{\emph{{On $B \to V \ell\ell$ at small
  dilepton invariant mass, power corrections, and new physics}},
  }{}\href{http://dx.doi.org/10.1007/JHEP05(2013)043}{JHEP \textbf{05} (2013)
  043}, \href{http://arxiv.org/abs/1212.2263}{{\normalfont\ttfamily
  arXiv:1212.2263}}\relax
\mciteBstWouldAddEndPuncttrue
\mciteSetBstMidEndSepPunct{\mcitedefaultmidpunct}
{\mcitedefaultendpunct}{\mcitedefaultseppunct}\relax
\EndOfBibitem
\bibitem{Beaujean:2013soa}
F.~Beaujean, C.~Bobeth, and D.~van Dyk,
  \ifthenelse{\boolean{articletitles}}{\emph{{Comprehensive Bayesian analysis
  of rare (semi)leptonic and radiative $B$ decays}},
  }{}\href{http://dx.doi.org/10.1140/epjc/s10052-014-2897-0}{Eur.\ Phys.\ J.\
  \textbf{C74} (2014) 2897},
  \href{http://arxiv.org/abs/1310.2478}{{\normalfont\ttfamily
  arXiv:1310.2478}}\relax
\mciteBstWouldAddEndPuncttrue
\mciteSetBstMidEndSepPunct{\mcitedefaultmidpunct}
{\mcitedefaultendpunct}{\mcitedefaultseppunct}\relax
\EndOfBibitem
\bibitem{Hurth:2013ssa}
T.~Hurth and F.~Mahmoudi, \ifthenelse{\boolean{articletitles}}{\emph{{On the
  LHCb anomaly in B $\to K^*\ell^+\ell^-$}},
  }{}\href{http://dx.doi.org/10.1007/JHEP04(2014)097}{JHEP \textbf{04} (2014)
  097}, \href{http://arxiv.org/abs/1312.5267}{{\normalfont\ttfamily
  arXiv:1312.5267}}\relax
\mciteBstWouldAddEndPuncttrue
\mciteSetBstMidEndSepPunct{\mcitedefaultmidpunct}
{\mcitedefaultendpunct}{\mcitedefaultseppunct}\relax
\EndOfBibitem
\bibitem{Gauld:2013qja}
R.~Gauld, F.~Goertz, and U.~Haisch,
  \ifthenelse{\boolean{articletitles}}{\emph{{An explicit $Z'$-boson
  explanation of the $B \to K^* \mu^+ \mu^-$ anomaly}},
  }{}\href{http://dx.doi.org/10.1007/JHEP01(2014)069}{JHEP \textbf{01} (2014)
  069}, \href{http://arxiv.org/abs/1310.1082}{{\normalfont\ttfamily
  arXiv:1310.1082}}\relax
\mciteBstWouldAddEndPuncttrue
\mciteSetBstMidEndSepPunct{\mcitedefaultmidpunct}
{\mcitedefaultendpunct}{\mcitedefaultseppunct}\relax
\EndOfBibitem
\bibitem{Datta:2013kja}
A.~Datta, M.~Duraisamy, and D.~Ghosh,
  \ifthenelse{\boolean{articletitles}}{\emph{{Explaining the $B \to K^\ast
  \mu^+ \mu^-$ data with scalar interactions}},
  }{}\href{http://dx.doi.org/10.1103/PhysRevD.89.071501}{Phys.\ Rev.\
  \textbf{D89} (2014) 071501},
  \href{http://arxiv.org/abs/1310.1937}{{\normalfont\ttfamily
  arXiv:1310.1937}}\relax
\mciteBstWouldAddEndPuncttrue
\mciteSetBstMidEndSepPunct{\mcitedefaultmidpunct}
{\mcitedefaultendpunct}{\mcitedefaultseppunct}\relax
\EndOfBibitem
\bibitem{Lyon:2014hpa}
J.~Lyon and R.~Zwicky, \ifthenelse{\boolean{articletitles}}{\emph{{Resonances
  gone topsy turvy - the charm of QCD or new physics in $b \to s \ell^+
  \ell^-$?}}, }{}\href{http://arxiv.org/abs/1406.0566}{{\normalfont\ttfamily
  arXiv:1406.0566}}\relax
\mciteBstWouldAddEndPuncttrue
\mciteSetBstMidEndSepPunct{\mcitedefaultmidpunct}
{\mcitedefaultendpunct}{\mcitedefaultseppunct}\relax
\EndOfBibitem
\bibitem{Descotes-Genon:2014uoa}
S.~Descotes-Genon, L.~Hofer, J.~Matias, and J.~Virto,
  \ifthenelse{\boolean{articletitles}}{\emph{{On the impact of power
  corrections in the prediction of $B \to K^*\mu^+\mu^-$ observables}},
  }{}\href{http://dx.doi.org/10.1007/JHEP12(2014)125}{JHEP \textbf{12} (2014)
  125}, \href{http://arxiv.org/abs/1407.8526}{{\normalfont\ttfamily
  arXiv:1407.8526}}\relax
\mciteBstWouldAddEndPuncttrue
\mciteSetBstMidEndSepPunct{\mcitedefaultmidpunct}
{\mcitedefaultendpunct}{\mcitedefaultseppunct}\relax
\EndOfBibitem
\bibitem{AltAndStraubLatest}
W.~Altmannshofer and D.~M. Straub,
  \ifthenelse{\boolean{articletitles}}{\emph{{New physics in $b\rightarrow s$
  transitions after LHC run 1}},
  }{}\href{http://dx.doi.org/10.1140/epjc/s10052-015-3602-7}{Eur.\ Phys.\ J.\
  \textbf{C75} (2015) 382},
  \href{http://arxiv.org/abs/1411.3161}{{\normalfont\ttfamily
  arXiv:1411.3161}}\relax
\mciteBstWouldAddEndPuncttrue
\mciteSetBstMidEndSepPunct{\mcitedefaultmidpunct}
{\mcitedefaultendpunct}{\mcitedefaultseppunct}\relax
\EndOfBibitem
\bibitem{Buttazzo:2016kid}
D.~Buttazzo, A.~Greljo, G.~Isidori, and D.~Marzocca,
  \ifthenelse{\boolean{articletitles}}{\emph{{Toward a coherent solution of
  diphoton and flavor anomalies}},
  }{}\href{http://dx.doi.org/10.1007/JHEP08(2016)035}{JHEP \textbf{08} (2016)
  035}, \href{http://arxiv.org/abs/1604.03940}{{\normalfont\ttfamily
  arXiv:1604.03940}}\relax
\mciteBstWouldAddEndPuncttrue
\mciteSetBstMidEndSepPunct{\mcitedefaultmidpunct}
{\mcitedefaultendpunct}{\mcitedefaultseppunct}\relax
\EndOfBibitem
\bibitem{Ciuchini:2015qxb}
M.~Ciuchini {\em et~al.}, \ifthenelse{\boolean{articletitles}}{\emph{{$B\to K^*
  \ell^+ \ell^-$ decays at large recoil in the Standard Model: A theoretical
  reappraisal}}, }{}\href{http://dx.doi.org/10.1007/JHEP06(2016)116}{JHEP
  \textbf{06} (2016) 116},
  \href{http://arxiv.org/abs/1512.07157}{{\normalfont\ttfamily
  arXiv:1512.07157}}\relax
\mciteBstWouldAddEndPuncttrue
\mciteSetBstMidEndSepPunct{\mcitedefaultmidpunct}
{\mcitedefaultendpunct}{\mcitedefaultseppunct}\relax
\EndOfBibitem
\bibitem{Lees:2012tva}
{BaBar collaboration}, J.~P. Lees {\em et~al.},
  \ifthenelse{\boolean{articletitles}}{\emph{{Measurement of branching
  fractions and rate asymmetries in the rare decays $B \to K^{(*)}
  \ellp\ellm$}}, }{}\href{http://dx.doi.org/10.1103/PhysRevD.86.032012}{Phys.\
  Rev.\  \textbf{D86} (2012) 032012},
  \href{http://arxiv.org/abs/1204.3933}{{\normalfont\ttfamily
  arXiv:1204.3933}}\relax
\mciteBstWouldAddEndPuncttrue
\mciteSetBstMidEndSepPunct{\mcitedefaultmidpunct}
{\mcitedefaultendpunct}{\mcitedefaultseppunct}\relax
\EndOfBibitem
\bibitem{Wei:2009zv}
\belle collaboration, J.-T. Wei {\em et~al.},
  \ifthenelse{\boolean{articletitles}}{\emph{{Measurement of the differential
  branching fraction and forward-backward asymmetry for $B \to
  K^{(*)}\ellp\ellm$}},
  }{}\href{http://dx.doi.org/10.1103/PhysRevLett.103.171801}{Phys.\ Rev.\
  Lett.\  \textbf{103} (2009) 171801},
  \href{http://arxiv.org/abs/0904.0770}{{\normalfont\ttfamily
  arXiv:0904.0770}}\relax
\mciteBstWouldAddEndPuncttrue
\mciteSetBstMidEndSepPunct{\mcitedefaultmidpunct}
{\mcitedefaultendpunct}{\mcitedefaultseppunct}\relax
\EndOfBibitem
\bibitem{Aaltonen:2011cn}
CDF collaboration, T.~Aaltonen {\em et~al.},
  \ifthenelse{\boolean{articletitles}}{\emph{{Measurement of the
  forward-backward asymmetry in the $B \to K^{(*)} \mu^+ \mu^-$ decay and first
  observation of the $B^0_s \to \phi \mu^+ \mu^-$ decay}},
  }{}\href{http://dx.doi.org/10.1103/PhysRevLett.106.161801}{Phys.\ Rev.\
  Lett.\  \textbf{106} (2011) 161801},
  \href{http://arxiv.org/abs/1101.1028}{{\normalfont\ttfamily
  arXiv:1101.1028}}\relax
\mciteBstWouldAddEndPuncttrue
\mciteSetBstMidEndSepPunct{\mcitedefaultmidpunct}
{\mcitedefaultendpunct}{\mcitedefaultseppunct}\relax
\EndOfBibitem
\bibitem{CMSKstmm}
CMS collaboration, V.~Khachatryan {\em et~al.},
  \ifthenelse{\boolean{articletitles}}{\emph{{Angular analysis of the decay
  $\Bd\to\Kstarz\mumu$ from $pp$ collisions at $\sqrt{s} = 8\tev$}},
  }{}\href{http://dx.doi.org/10.1016/j.physletb.2015.12.020}{Phys.\ Lett.\
  \textbf{B753} (2016) 424},
  \href{http://arxiv.org/abs/1507.08126}{{\normalfont\ttfamily
  arXiv:1507.08126}}\relax
\mciteBstWouldAddEndPuncttrue
\mciteSetBstMidEndSepPunct{\mcitedefaultmidpunct}
{\mcitedefaultendpunct}{\mcitedefaultseppunct}\relax
\EndOfBibitem
\bibitem{LHCb-PAPER-2014-006}
LHCb collaboration, R.~Aaij {\em et~al.},
  \ifthenelse{\boolean{articletitles}}{\emph{{Differential branching fractions
  and isospin asymmetries of $B \to K^{(*)}\mu^+\mu^-$ decays}},
  }{}\href{http://dx.doi.org/10.1007/JHEP06(2014)133}{JHEP \textbf{06} (2014)
  133}, \href{http://arxiv.org/abs/1403.8044}{{\normalfont\ttfamily
  arXiv:1403.8044}}\relax
\mciteBstWouldAddEndPuncttrue
\mciteSetBstMidEndSepPunct{\mcitedefaultmidpunct}
{\mcitedefaultendpunct}{\mcitedefaultseppunct}\relax
\EndOfBibitem
\bibitem{LHCb-PAPER-2015-051}
LHCb collaboration, R.~Aaij {\em et~al.},
  \ifthenelse{\boolean{articletitles}}{\emph{{Angular analysis of the $B^{0}
  \to K^{*0} \mu^{+} \mu^{-}$ decay using 3 fb$^{-1}$ of integrated
  luminosity}}, }{}\href{http://dx.doi.org/10.1007/JHEP02(2016)104}{JHEP
  \textbf{02} (2016) 104},
  \href{http://arxiv.org/abs/1512.04442}{{\normalfont\ttfamily
  arXiv:1512.04442}}\relax
\mciteBstWouldAddEndPuncttrue
\mciteSetBstMidEndSepPunct{\mcitedefaultmidpunct}
{\mcitedefaultendpunct}{\mcitedefaultseppunct}\relax
\EndOfBibitem
\bibitem{Kruger:1996cv}
F.~Kruger and L.~M. Sehgal, \ifthenelse{\boolean{articletitles}}{\emph{{Lepton
  polarization in the decays $B\to X_s \mu^+\mu^-$ and $B\to X_s \tau^+
  \tau^-$}}, }{}\href{http://dx.doi.org/10.1016/0370-2693(96)00413-3}{Phys.\
  Lett.\  \textbf{B380} (1996) 199},
  \href{http://arxiv.org/abs/hep-ph/9603237}{{\normalfont\ttfamily
  arXiv:hep-ph/9603237}}\relax
\mciteBstWouldAddEndPuncttrue
\mciteSetBstMidEndSepPunct{\mcitedefaultmidpunct}
{\mcitedefaultendpunct}{\mcitedefaultseppunct}\relax
\EndOfBibitem
\bibitem{Bai:2001ct}
BES collaboration, J.~Z. Bai {\em et~al.},
  \ifthenelse{\boolean{articletitles}}{\emph{{Measurements of the cross-section
  for $\epem \to$ hadrons at center-of-mass energies from 2\gev to 5\gev}},
  }{}\href{http://dx.doi.org/10.1103/PhysRevLett.88.101802}{Phys.\ Rev.\ Lett.\
   \textbf{88} (2002) 101802},
  \href{http://arxiv.org/abs/hep-ex/0102003}{{\normalfont\ttfamily
  arXiv:hep-ex/0102003}}\relax
\mciteBstWouldAddEndPuncttrue
\mciteSetBstMidEndSepPunct{\mcitedefaultmidpunct}
{\mcitedefaultendpunct}{\mcitedefaultseppunct}\relax
\EndOfBibitem
\bibitem{LHCb-PAPER-2013-039}
LHCb collaboration, R.~Aaij {\em et~al.},
  \ifthenelse{\boolean{articletitles}}{\emph{{Observation of a resonance in
  $B^+\to K^+\mu^+\mu^-$ decays at low recoil}},
  }{}\href{http://dx.doi.org/10.1103/PhysRevLett.111.112003}{Phys.\ Rev.\
  Lett.\  \textbf{111} (2013) 112003},
  \href{http://arxiv.org/abs/1307.7595}{{\normalfont\ttfamily
  arXiv:1307.7595}}\relax
\mciteBstWouldAddEndPuncttrue
\mciteSetBstMidEndSepPunct{\mcitedefaultmidpunct}
{\mcitedefaultendpunct}{\mcitedefaultseppunct}\relax
\EndOfBibitem
\bibitem{Alves:2008zz}
LHCb collaboration, A.~A. Alves~Jr.\ {\em et~al.},
  \ifthenelse{\boolean{articletitles}}{\emph{{The \lhcb detector at the LHC}},
  }{}\href{http://dx.doi.org/10.1088/1748-0221/3/08/S08005}{JINST \textbf{3}
  (2008) S08005}\relax
\mciteBstWouldAddEndPuncttrue
\mciteSetBstMidEndSepPunct{\mcitedefaultmidpunct}
{\mcitedefaultendpunct}{\mcitedefaultseppunct}\relax
\EndOfBibitem
\bibitem{LHCb-DP-2014-002}
LHCb collaboration, R.~Aaij {\em et~al.},
  \ifthenelse{\boolean{articletitles}}{\emph{{LHCb detector performance}},
  }{}\href{http://dx.doi.org/10.1142/S0217751X15300227}{Int.\ J.\ Mod.\ Phys.\
  \textbf{A30} (2015) 1530022},
  \href{http://arxiv.org/abs/1412.6352}{{\normalfont\ttfamily
  arXiv:1412.6352}}\relax
\mciteBstWouldAddEndPuncttrue
\mciteSetBstMidEndSepPunct{\mcitedefaultmidpunct}
{\mcitedefaultendpunct}{\mcitedefaultseppunct}\relax
\EndOfBibitem
\bibitem{LHCb-PAPER-2012-048}
LHCb collaboration, R.~Aaij {\em et~al.},
  \ifthenelse{\boolean{articletitles}}{\emph{{Measurements of the
  $\Lambda_b^0$, $\Xi_b^-$, and $\Omega_b^-$ baryon masses}},
  }{}\href{http://dx.doi.org/10.1103/PhysRevLett.110.182001}{Phys.\ Rev.\
  Lett.\  \textbf{110} (2013) 182001},
  \href{http://arxiv.org/abs/1302.1072}{{\normalfont\ttfamily
  arXiv:1302.1072}}\relax
\mciteBstWouldAddEndPuncttrue
\mciteSetBstMidEndSepPunct{\mcitedefaultmidpunct}
{\mcitedefaultendpunct}{\mcitedefaultseppunct}\relax
\EndOfBibitem
\bibitem{LHCb-DP-2012-004}
R.~Aaij {\em et~al.}, \ifthenelse{\boolean{articletitles}}{\emph{{The \lhcb
  trigger and its performance in 2011}},
  }{}\href{http://dx.doi.org/10.1088/1748-0221/8/04/P04022}{JINST \textbf{8}
  (2013) P04022}, \href{http://arxiv.org/abs/1211.3055}{{\normalfont\ttfamily
  arXiv:1211.3055}}\relax
\mciteBstWouldAddEndPuncttrue
\mciteSetBstMidEndSepPunct{\mcitedefaultmidpunct}
{\mcitedefaultendpunct}{\mcitedefaultseppunct}\relax
\EndOfBibitem
\bibitem{Sjostrand:2006za}
T.~Sj\"{o}strand, S.~Mrenna, and P.~Skands,
  \ifthenelse{\boolean{articletitles}}{\emph{{PYTHIA 6.4 physics and manual}},
  }{}\href{http://dx.doi.org/10.1088/1126-6708/2006/05/026}{JHEP \textbf{05}
  (2006) 026}, \href{http://arxiv.org/abs/hep-ph/0603175}{{\normalfont\ttfamily
  arXiv:hep-ph/0603175}}\relax
\mciteBstWouldAddEndPuncttrue
\mciteSetBstMidEndSepPunct{\mcitedefaultmidpunct}
{\mcitedefaultendpunct}{\mcitedefaultseppunct}\relax
\EndOfBibitem
\bibitem{Sjostrand:2007gs}
T.~Sj\"{o}strand, S.~Mrenna, and P.~Skands,
  \ifthenelse{\boolean{articletitles}}{\emph{{A brief introduction to PYTHIA
  8.1}}, }{}\href{http://dx.doi.org/10.1016/j.cpc.2008.01.036}{Comput.\ Phys.\
  Commun.\  \textbf{178} (2008) 852},
  \href{http://arxiv.org/abs/0710.3820}{{\normalfont\ttfamily
  arXiv:0710.3820}}\relax
\mciteBstWouldAddEndPuncttrue
\mciteSetBstMidEndSepPunct{\mcitedefaultmidpunct}
{\mcitedefaultendpunct}{\mcitedefaultseppunct}\relax
\EndOfBibitem
\bibitem{LHCb-PROC-2010-056}
I.~Belyaev {\em et~al.}, \ifthenelse{\boolean{articletitles}}{\emph{{Handling
  of the generation of primary events in Gauss, the LHCb simulation
  framework}}, }{}\href{http://dx.doi.org/10.1088/1742-6596/331/3/032047}{{J.\
  Phys.\ Conf.\ Ser.\ } \textbf{331} (2011) 032047}\relax
\mciteBstWouldAddEndPuncttrue
\mciteSetBstMidEndSepPunct{\mcitedefaultmidpunct}
{\mcitedefaultendpunct}{\mcitedefaultseppunct}\relax
\EndOfBibitem
\bibitem{Lange:2001uf}
D.~J. Lange, \ifthenelse{\boolean{articletitles}}{\emph{{The EvtGen particle
  decay simulation package}},
  }{}\href{http://dx.doi.org/10.1016/S0168-9002(01)00089-4}{Nucl.\ Instrum.\
  Meth.\  \textbf{A462} (2001) 152}\relax
\mciteBstWouldAddEndPuncttrue
\mciteSetBstMidEndSepPunct{\mcitedefaultmidpunct}
{\mcitedefaultendpunct}{\mcitedefaultseppunct}\relax
\EndOfBibitem
\bibitem{Golonka:2005pn}
P.~Golonka and Z.~Was, \ifthenelse{\boolean{articletitles}}{\emph{{PHOTOS Monte
  Carlo: A precision tool for QED corrections in $Z$ and $W$ decays}},
  }{}\href{http://dx.doi.org/10.1140/epjc/s2005-02396-4}{Eur.\ Phys.\ J.\
  \textbf{C45} (2006) 97},
  \href{http://arxiv.org/abs/hep-ph/0506026}{{\normalfont\ttfamily
  arXiv:hep-ph/0506026}}\relax
\mciteBstWouldAddEndPuncttrue
\mciteSetBstMidEndSepPunct{\mcitedefaultmidpunct}
{\mcitedefaultendpunct}{\mcitedefaultseppunct}\relax
\EndOfBibitem
\bibitem{LHCb-PROC-2011-006}
M.~Clemencic {\em et~al.}, \ifthenelse{\boolean{articletitles}}{\emph{{The
  \lhcb simulation application, Gauss: Design, evolution and experience}},
  }{}\href{http://dx.doi.org/10.1088/1742-6596/331/3/032023}{{J.\ Phys.\ Conf.\
  Ser.\ } \textbf{331} (2011) 032023}\relax
\mciteBstWouldAddEndPuncttrue
\mciteSetBstMidEndSepPunct{\mcitedefaultmidpunct}
{\mcitedefaultendpunct}{\mcitedefaultseppunct}\relax
\EndOfBibitem
\bibitem{Allison:2006ve}
Geant4 collaboration, J.~Allison {\em et~al.},
  \ifthenelse{\boolean{articletitles}}{\emph{{Geant4 developments and
  applications}}, }{}\href{http://dx.doi.org/10.1109/TNS.2006.869826}{IEEE
  Trans.\ Nucl.\ Sci.\  \textbf{53} (2006) 270}\relax
\mciteBstWouldAddEndPuncttrue
\mciteSetBstMidEndSepPunct{\mcitedefaultmidpunct}
{\mcitedefaultendpunct}{\mcitedefaultseppunct}\relax
\EndOfBibitem
\bibitem{Agostinelli:2002hh}
Geant4 collaboration, S.~Agostinelli {\em et~al.},
  \ifthenelse{\boolean{articletitles}}{\emph{{Geant4: A simulation toolkit}},
  }{}\href{http://dx.doi.org/10.1016/S0168-9002(03)01368-8}{Nucl.\ Instrum.\
  Meth.\  \textbf{A506} (2003) 250}\relax
\mciteBstWouldAddEndPuncttrue
\mciteSetBstMidEndSepPunct{\mcitedefaultmidpunct}
{\mcitedefaultendpunct}{\mcitedefaultseppunct}\relax
\EndOfBibitem
\bibitem{Breiman}
L.~Breiman, J.~H. Friedman, R.~A. Olshen, and C.~J. Stone, {\em Classification
  and regression trees}, Wadsworth international group, Belmont, California,
  USA, 1984\relax
\mciteBstWouldAddEndPuncttrue
\mciteSetBstMidEndSepPunct{\mcitedefaultmidpunct}
{\mcitedefaultendpunct}{\mcitedefaultseppunct}\relax
\EndOfBibitem
\bibitem{AdaBoost}
Y.~Freund and R.~E. Schapire, \ifthenelse{\boolean{articletitles}}{\emph{A
  decision-theoretic generalization of on-line learning and an application to
  boosting}, }{}\href{http://dx.doi.org/10.1006/jcss.1997.1504}{J.\ Comput.\
  and Syst.\ Sci.\  \textbf{55} (1997) 119}\relax
\mciteBstWouldAddEndPuncttrue
\mciteSetBstMidEndSepPunct{\mcitedefaultmidpunct}
{\mcitedefaultendpunct}{\mcitedefaultseppunct}\relax
\EndOfBibitem
\bibitem{Olive:2016xmw}
Particle Data Group, C.~Patrignani {\em et~al.},
  \ifthenelse{\boolean{articletitles}}{\emph{{Review of Particle Physics}},
  }{}\href{http://dx.doi.org/10.1088/1674-1137/40/10/100001}{Chin.\ Phys.\
  \textbf{C40} (2016) 100001}\relax
\mciteBstWouldAddEndPuncttrue
\mciteSetBstMidEndSepPunct{\mcitedefaultmidpunct}
{\mcitedefaultendpunct}{\mcitedefaultseppunct}\relax
\EndOfBibitem
\bibitem{Bailey:2015dka}
J.~A. Bailey {\em et~al.}, \ifthenelse{\boolean{articletitles}}{\emph{{$B\to
  Kl^+l^-$ decay form factors from three-flavor lattice QCD}},
  }{}\href{http://dx.doi.org/10.1103/PhysRevD.93.025026}{Phys.\ Rev.\
  \textbf{D93} (2016) 025026},
  \href{http://arxiv.org/abs/1509.06235}{{\normalfont\ttfamily
  arXiv:1509.06235}}\relax
\mciteBstWouldAddEndPuncttrue
\mciteSetBstMidEndSepPunct{\mcitedefaultmidpunct}
{\mcitedefaultendpunct}{\mcitedefaultseppunct}\relax
\EndOfBibitem
\bibitem{Altmannshofer:2008dz}
W.~Altmannshofer {\em et~al.},
  \ifthenelse{\boolean{articletitles}}{\emph{{Symmetries and asymmetries of $B
  \to K^{*} \mu^{+} \mu^{-}$ decays in the Standard Model and beyond}},
  }{}\href{http://dx.doi.org/10.1088/1126-6708/2009/01/019}{JHEP \textbf{01}
  (2009) 019}, \href{http://arxiv.org/abs/0811.1214}{{\normalfont\ttfamily
  arXiv:0811.1214}}\relax
\mciteBstWouldAddEndPuncttrue
\mciteSetBstMidEndSepPunct{\mcitedefaultmidpunct}
{\mcitedefaultendpunct}{\mcitedefaultseppunct}\relax
\EndOfBibitem
\bibitem{Alok:2011gv}
A.~K. Alok {\em et~al.}, \ifthenelse{\boolean{articletitles}}{\emph{{New
  physics in $b \to s \mu^+ \mu^-$: $C\!P$-violating observables}},
  }{}\href{http://dx.doi.org/10.1007/JHEP11(2011)122}{JHEP \textbf{11} (2011)
  122}, \href{http://arxiv.org/abs/1103.5344}{{\normalfont\ttfamily
  arXiv:1103.5344}}\relax
\mciteBstWouldAddEndPuncttrue
\mciteSetBstMidEndSepPunct{\mcitedefaultmidpunct}
{\mcitedefaultendpunct}{\mcitedefaultseppunct}\relax
\EndOfBibitem
\bibitem{LHCb-PAPER-2014-032}
LHCb collaboration, R.~Aaij {\em et~al.},
  \ifthenelse{\boolean{articletitles}}{\emph{{Measurement of $C\!P$ asymmetries
  in the decays $B^0 \to K^{*0}\mu^+\mu^-$ and $B^+ \to K^+\mu^+\mu^-$}},
  }{}\href{http://dx.doi.org/10.1007/JHEP09(2014)177}{JHEP \textbf{09} (2014)
  177}, \href{http://arxiv.org/abs/1408.0978}{{\normalfont\ttfamily
  arXiv:1408.0978}}\relax
\mciteBstWouldAddEndPuncttrue
\mciteSetBstMidEndSepPunct{\mcitedefaultmidpunct}
{\mcitedefaultendpunct}{\mcitedefaultseppunct}\relax
\EndOfBibitem
\bibitem{Feldmann:2002iw}
T.~Feldmann and J.~Matias, \ifthenelse{\boolean{articletitles}}{\emph{{Forward
  backward and isospin asymmetry for $B \to K^* l^+ l^-$ decay in the Standard
  Model and in supersymmetry}},
  }{}\href{http://dx.doi.org/10.1088/1126-6708/2003/01/074}{JHEP \textbf{01}
  (2003) 074}, \href{http://arxiv.org/abs/hep-ph/0212158}{{\normalfont\ttfamily
  arXiv:hep-ph/0212158}}\relax
\mciteBstWouldAddEndPuncttrue
\mciteSetBstMidEndSepPunct{\mcitedefaultmidpunct}
{\mcitedefaultendpunct}{\mcitedefaultseppunct}\relax
\EndOfBibitem
\bibitem{Khodjamirian:2012rm}
A.~Khodjamirian, T.~Mannel, and Y.~M. Wang,
  \ifthenelse{\boolean{articletitles}}{\emph{{$B \to K \ell^{+}\ell^{-}$ decay
  at large hadronic recoil}},
  }{}\href{http://dx.doi.org/10.1007/JHEP02(2013)010}{JHEP \textbf{02} (2013)
  010}, \href{http://arxiv.org/abs/1211.0234}{{\normalfont\ttfamily
  arXiv:1211.0234}}\relax
\mciteBstWouldAddEndPuncttrue
\mciteSetBstMidEndSepPunct{\mcitedefaultmidpunct}
{\mcitedefaultendpunct}{\mcitedefaultseppunct}\relax
\EndOfBibitem
\bibitem{Lyon:2013gba}
J.~Lyon and R.~Zwicky, \ifthenelse{\boolean{articletitles}}{\emph{{Isospin
  asymmetries in $B\to(K^*,\rho)\gamma/\ell^+\ell^-$ and $B\to K\ell^+\ell^-$
  in and beyond the Standard Model}},
  }{}\href{http://dx.doi.org/10.1103/PhysRevD.88.094004}{Phys.\ Rev.\
  \textbf{D88} (2013) 094004},
  \href{http://arxiv.org/abs/1305.4797}{{\normalfont\ttfamily
  arXiv:1305.4797}}\relax
\mciteBstWouldAddEndPuncttrue
\mciteSetBstMidEndSepPunct{\mcitedefaultmidpunct}
{\mcitedefaultendpunct}{\mcitedefaultseppunct}\relax
\EndOfBibitem
\bibitem{Flatte:1976xu}
S.~M. Flatt{\'e}, \ifthenelse{\boolean{articletitles}}{\emph{{Coupled-channel
  analysis of the $\pi\eta$ and $K\overline{K}$ systems near $K\overline{K}$
  threshold}}, }{}\href{http://dx.doi.org/10.1016/0370-2693(76)90654-7}{Phys.\
  Lett.\  \textbf{B63} (1976) 224}\relax
\mciteBstWouldAddEndPuncttrue
\mciteSetBstMidEndSepPunct{\mcitedefaultmidpunct}
{\mcitedefaultendpunct}{\mcitedefaultseppunct}\relax
\EndOfBibitem
\bibitem{Bourrely:2008za}
C.~Bourrely, I.~Caprini, and L.~Lellouch,
  \ifthenelse{\boolean{articletitles}}{\emph{{Model-independent description of
  $B \to \pi l \nu$ decays and a determination of $|V_{ub}|$}},
  }{}\href{http://dx.doi.org/10.1103/PhysRevD.79.013008}{Phys.\ Rev.\
  \textbf{D79} (2009) 013008},
  \href{http://arxiv.org/abs/0807.2722}{{\normalfont\ttfamily
  arXiv:0807.2722}},
  \href{http://dx.doi.org/10.1103/PhysRevD.82.099902}{[Erratum: Phys.
  Rev.\,{\bf D82} (2010) 099902]}\relax
\mciteBstWouldAddEndPuncttrue
\mciteSetBstMidEndSepPunct{\mcitedefaultmidpunct}
{\mcitedefaultendpunct}{\mcitedefaultseppunct}\relax
\EndOfBibitem
\bibitem{Ablikim:2007gd}
BES collaboration, M.~Ablikim {\em et~al.},
  \ifthenelse{\boolean{articletitles}}{\emph{{Determination of the
  $\psi(3770)$, $\psi(4040)$, $\psi(4160)$ and $\psi(4415)$ resonance
  parameters}},
  }{}\href{http://dx.doi.org/10.1016/j.physletb.2007.11.100}{Phys.\ Lett.\
  \textbf{B660} (2008) 315},
  \href{http://arxiv.org/abs/0705.4500}{{\normalfont\ttfamily
  arXiv:0705.4500}}\relax
\mciteBstWouldAddEndPuncttrue
\mciteSetBstMidEndSepPunct{\mcitedefaultmidpunct}
{\mcitedefaultendpunct}{\mcitedefaultseppunct}\relax
\EndOfBibitem
\bibitem{Jung:2015yma}
M.~Jung, \ifthenelse{\boolean{articletitles}}{\emph{{Branching ratio
  measurements and isospin violation in B-meson decays}},
  }{}\href{http://dx.doi.org/10.1016/j.physletb.2015.12.024}{Phys.\ Lett.\
  \textbf{B753} (2016) 187},
  \href{http://arxiv.org/abs/1510.03423}{{\normalfont\ttfamily
  arXiv:1510.03423}}\relax
\mciteBstWouldAddEndPuncttrue
\mciteSetBstMidEndSepPunct{\mcitedefaultmidpunct}
{\mcitedefaultendpunct}{\mcitedefaultseppunct}\relax
\EndOfBibitem
\bibitem{Cooley:1965zz}
J.~W. Cooley and J.~W. Tukey, \ifthenelse{\boolean{articletitles}}{\emph{{An
  algorithm for the machine calculation of complex Fourier series}},
  }{}\href{http://dx.doi.org/10.1090/S0025-5718-1965-0178586-1}{Math.\ Comput.\
   \textbf{19} (1965) 297}\relax
\mciteBstWouldAddEndPuncttrue
\mciteSetBstMidEndSepPunct{\mcitedefaultmidpunct}
{\mcitedefaultendpunct}{\mcitedefaultseppunct}\relax
\EndOfBibitem
\bibitem{FFTW05}
M.~Frigo and S.~G. Johnson, \ifthenelse{\boolean{articletitles}}{\emph{The
  design and implementation of {FFTW3}},
  }{}\href{http://dx.doi.org/10.1109/JPROC.2004.840301}{Proceedings of the IEEE
  \textbf{93} (2005) 216}\relax
\mciteBstWouldAddEndPuncttrue
\mciteSetBstMidEndSepPunct{\mcitedefaultmidpunct}
{\mcitedefaultendpunct}{\mcitedefaultseppunct}\relax
\EndOfBibitem
\bibitem{Albrecht:1994tb}
ARGUS collaboration, H.~Albrecht {\em et~al.},
  \ifthenelse{\boolean{articletitles}}{\emph{{Measurement of the polarization
  in the decay $B \to \jpsi \Kstar$}},
  }{}\href{http://dx.doi.org/10.1016/0370-2693(94)01302-0}{Phys.\ Lett.\
  \textbf{B340} (1994) 217}\relax
\mciteBstWouldAddEndPuncttrue
\mciteSetBstMidEndSepPunct{\mcitedefaultmidpunct}
{\mcitedefaultendpunct}{\mcitedefaultseppunct}\relax
\EndOfBibitem
\end{mcitethebibliography}

\ifx\mcitethebibliography\mciteundefinedmacro
\PackageError{LHCb.bst}{mciteplus.sty has not been loaded}
{This bibstyle requires the use of the mciteplus package.}\fi
\providecommand{\href}[2]{#2}

\newpage

\newpage
\centerline{\large\bf LHCb collaboration}
\begin{flushleft}
\small
R.~Aaij$^{40}$,
B.~Adeva$^{39}$,
M.~Adinolfi$^{48}$,
Z.~Ajaltouni$^{5}$,
S.~Akar$^{59}$,
J.~Albrecht$^{10}$,
F.~Alessio$^{40}$,
M.~Alexander$^{53}$,
S.~Ali$^{43}$,
G.~Alkhazov$^{31}$
% ...

{\footnotesize \it
% ...
$ ^{5}$Clermont Universit{\'e}, Universit{\'e} Blaise Pascal, CNRS/IN2P3, LPC, Clermont-Ferrand, France\\
% ...
$ ^{10}$Fakult{\"a}t Physik, Technische Universit{\"a}t Dortmund, Dortmund, Germany\\
% ...
$ ^{31}$Petersburg Nuclear Physics Institute (PNPI), Gatchina, Russia\\
% ...
$ ^{39}$Universidad de Santiago de Compostela, Santiago de Compostela, Spain\\
$ ^{40}$European Organization for Nuclear Research (CERN), Geneva, Switzerland\\
% ...
$ ^{43}$Nikhef National Institute for Subatomic Physics, Amsterdam, The Netherlands\\
% ...
$ ^{48}$H.H. Wills Physics Laboratory, University of Bristol, Bristol, United Kingdom\\
% ...
$ ^{53}$School of Physics and Astronomy, University of Glasgow, Glasgow, United Kingdom\\
% ...
$ ^{59}$University of Cincinnati, Cincinnati, OH, United States\\
% ...
}
\end{flushleft}

\end{document}